\documentclass[letter, amsfonts, amssymb, amsmath, reprint, showkeys, 
twoside, twocolumn,
elsarticle,
superscriptaddress]{revtex4-1}

\usepackage{xcolor}
\usepackage{graphicx} 
\usepackage{subfigure}
\usepackage{dcolumn} 
\usepackage{bm}
\usepackage{float}
\usepackage[columnwise,running]{lineno}
\usepackage[normalem]{ulem}
\usepackage{hyperref}
\usepackage{color}
\definecolor{orange}{cmyk}{0.,0.353,1.,0.}    % orange

\usepackage{tikz}
\usetikzlibrary{positioning}

\begin{document}

\title{Event Shape Selection Method in Search of the Chiral Magnetic Effect in Heavy-ion Collisions}

\author{Zhiwan Xu}\email{zhiwanxu@physics.ucla.edu}
\affiliation{Department of Physics and Astronomy, University of
  California, Los Angeles, California 90095, USA}
  
\author{Brian Chan}
\affiliation{Department of Physics and Astronomy, University of
  California, Los Angeles, California 90095, USA}
  
  \author{Gang Wang}\email{gwang@physics.ucla.edu}
\affiliation{Department of Physics and Astronomy, University of
  California, Los Angeles, California 90095, USA} 
\author{Aihong Tang} \affiliation{Brookhaven National Laboratory, Upton, New York 11973, USA} 
  \author{Huan Zhong Huang} \affiliation{Department of Physics and
  Astronomy, University of California, Los Angeles, California 90095,
  USA} \affiliation{Key Laboratory of Nuclear Physics and Ion-beam
  Application (MOE), and Institute of Modern Physics, Fudan
  University, Shanghai-200433, People’s Republic of China}
  %\date{February 2023}

\begin{abstract}
The search for the chiral magnetic effect (CME) in heavy-ion collisions has been impeded by the significant background arising from the anisotropic particle emission pattern, particularly elliptic flow.
To alleviate this background, the event shape selection (ESS) technique categorizes collision events according to their shapes and projects the CME observables to a class of events with minimal flow. 
In this study, we explore two event shape variables to classify events and two elliptic flow variables to
regulate the background.
Each type of variable can be calculated from either single particles or particle pairs, resulting in four combinations of event shape and elliptic flow variables.
By employing a toy model and the realistic event generator, event-by-event anomalous-viscous fluid dynamics (EBE-AVFD), we discover that the elliptic flow of resonances exhibits correlations with both the background and the potential CME signal, making the resonance flow unsuitable for background control.
Through the EBE-AVFD simulations of Au+Au collisions at $\sqrt{s_{NN}} = 200$ GeV with various input scenarios, we ascertain that the optimal ESS
strategy for background control entails utilizing the single-particle elliptic flow in conjunction with the event shape variable based on particle pairs.
\begin{description}
\item[keywords]
event shape; heavy-ion collision; chiral magnetic effect
\end{description}
\end{abstract}
\maketitle

\section{Introduction} \label{intro}

Ultra-relativistic heavy-ion collisions offer an opportunity to study the topological sector of quantum chromodynamics through the creation of a deconfined quark-gluon plasma. 
The deconfinement is accompanied by chiral symmetry restoration, whereby light quarks become nearly massless and acquire definite chirality.
Quarks with a chirality imbalance can experience an electric charge separation along an intense magnetic field ($\vec{B}$), known as the chiral magnetic effect (CME)~\cite{CME-10,CME-4,CME-5,CME-6}. 
The chirality imbalance may stem from the quantum chiral anomaly during topological vacuum transitions,
which violates parity and charge-parity symmetries in the strong interaction~\cite{CME-1,CME-2,CME-3}.
Intense magnetic fields on the order of $10^{18}$ Gauss~\cite{CME-7,CME-8}  can emerge as
the fragments of the colliding nuclei fly by each other in the experiments at the BNL Relativistic Heavy Ion Collider (RHIC) and the CERN Large Ion Collider (LHC).
Over the past two decades,  the CME has stimulated great interest in experimental investigations at RHIC and the LHC, where all its
preconditions could be satisfied.

\begin{figure}[tbhp]
\centering
\includegraphics[scale=0.42]{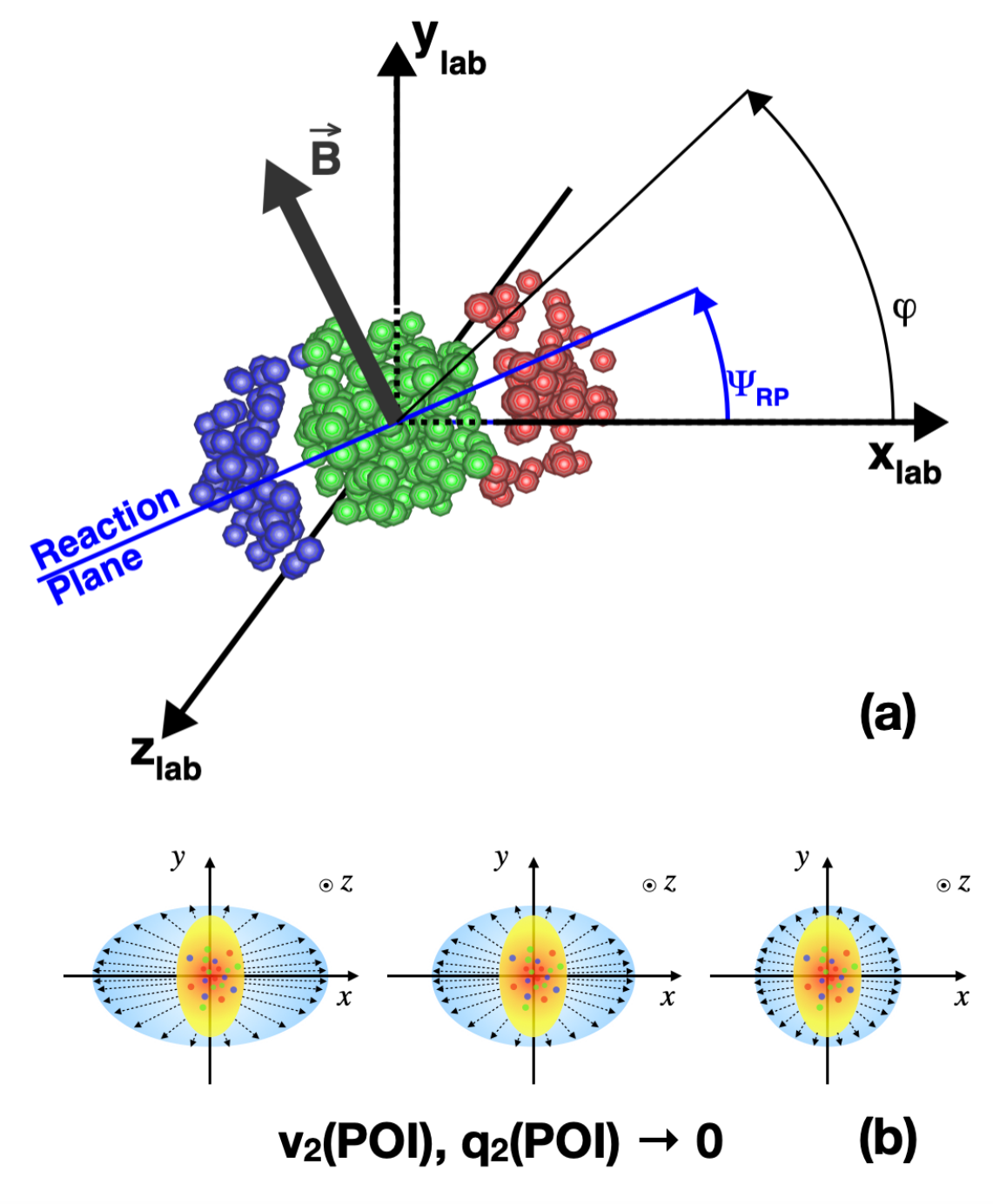}
\caption{(a) Sketch of a two-nucleus collision in the lab frame, with the left one exiting and the right one entering the page. Particle production occurs
in the overlap region. The
azimuthal angles of the reaction plane and a produced particle
 are illustrated.
 The magnetic field is perpendicular to the reaction plane.
(b) Schematic diagram of the event shape selection based on the particle emission pattern in the transverse plane. Both $v_2$ and $q_2$ for particles of interest could fluctuate towards zero despite a finite eccentricity of the overlap region.}
\label{setup}
\end{figure}

On average, the direction of ${\vec B}$ is perpendicular to the reaction plane (${\rm \Psi_{RP}}$), which is defined by impact parameter and beam momenta as illustrated in Fig.~\ref{setup}(a). Consequently, the CME leads to a separation of electric charges normal to the reaction plane.
To quantify the CME-induced charge separation and other collectivity modes, we adopt the convention of expressing the azimuthal angle distribution of final-state particles as a Fourier series~\cite{Sergeiflow1}
\begin{equation}
\frac{dN_{\alpha}}{d\varphi} \propto 1+ 2a_{1,\alpha}\sin\Delta\varphi + \sum_{n=1}^\infty 2v_{n,\alpha}\cos n\Delta\varphi , 
\label{equ:Fourier_expansion}
\end{equation}
where $\Delta\varphi = \varphi-\Psi_{\rm{RP}}$,  $\varphi$ represents the azimuthal angle of a particle's momentum, and $\alpha$ denotes its charge sign. 
$a_{1,+}$ and $a_{1,-}$  should bear 
opposite signs due to the charge separation, 
and they could have slightly different magnitudes in the presence of extra positive charges in the collision system.
The hydrodynamic expansion~\cite{hydro} supposedly transforms the anisotropic initial state in coordinate space into the anisotropy ($v_n$) of final-state particles in momentum space.
Of all the $v_n$ coefficients, the most relevant to the CME search is $v_2\equiv \langle\cos2\Delta\varphi\rangle$, elliptic flow.
When averaged over events, $v_2$ is roughly proportional to the  eccentricity of the initial overlap region. 
Later, we will explain the relationship between $v_2$ and the background in the CME observable, as well as how to suppress such contributions.

A recent study~\cite{AVFD5} has confirmed the similarity in the core components of the CME observables, such as the $\gamma_{112}$ correlator~\cite{CME-9}, the $R_{\Psi_2}(\Delta S)$ correlator~\cite{Magdy:2017yje}, and the signed balance functions~\cite{Tang:2019pbl}.
In this paper, we focus on the most widely used observable~\cite{CME-9},
\begin{equation}
\gamma_{112} \equiv \langle \cos(\varphi_\alpha + \varphi_\beta - 2\Psi_{\rm{RP}})\rangle,
\end{equation}
which is averaged over particle pairs and over events.
The difference between opposite-sign (OS) and same-sign (SS) pairs contains the true charge separation signal,
\begin{eqnarray}
\Delta\gamma_{112} &\equiv& \gamma^{\rm{OS}}_{112}- \gamma^{\rm{SS}}_{112} \\
&=&(a_{1,+}^2+a_{1,-}^2)/2-a_{1,+}a_{1,-} + {\rm BG} \\
&=& 2a_1^2 + {\rm BG}, \label{eq:dg2}
\end{eqnarray}
where $2a_1^2 \equiv (a_{1,+}^2+a_{1,-}^2)/2-a_{1,+}a_{1,-}$ is shorthand for the CME signal, and BG denotes background contributions to be reckoned with.
Experiments at RHIC and the LHC, including STAR~\cite{STAR-1,STAR-2,STAR-3,LPV_STAR5,STAR-4,STAR-5,STAR-6,STAR-7,STAR-8,STAR-9}, ALICE~\cite{ALICE-1,ALICE-2,ALICE-3}, and CMS~\cite{CMS-1,CMS-2}, have reported positively finite $\Delta\gamma_{112}$ measurements for various collision systems and beam energies. However, the data interpretation is hindered by an incomplete understanding of the background.
For instance, hitherto, the recent isobar data from STAR~\cite{STAR-7} cannot yield a definite conclusion on the existence of the CME  because of the overwhelming background in addition to the multiplicity mismatch between the two isobaric systems.

The BG term in Eq.~(\ref{eq:dg2}) consists of both flow and nonflow contributions. Nonflow correlations are unrelated to the reaction plane and originate from various sources such as clusters, resonances, jets, or di-jets. These mechanisms can simultaneously contaminate $\varphi_\alpha$, $\varphi_\beta$, and the estimation of $\Psi_{\rm RP}$, leading to a finite $\Delta\gamma_{112}$ value even in the absence of genuine $a_1$.
Nonflow effects are more conspicuous in smaller systems with lower multiplicities, like $p$+Au and $d$+Au collisions at $\sqrt{s_{NN}} = 200$ GeV~\cite{STAR-5}. 
For the simulation study presented in this work, we will exclusively employ the true reaction plane to avoid nonflow contamination.
In practical applications, the reaction plane can be approximated with the event plane defined by spectator nucleons to eliminate nonflow~\cite{STAR-8,STAR-9}.
As spectators exit the participant zone early in the collision, the influence of most nonflow mechanisms becomes negligible when the spectator plane is used. While the momentum conservation effect could potentially impact spectators, deploying two spectator planes symmetrically positioned in rapidity effectively eliminates this effect, as demonstrated in prior flow measurements~\cite{v1_STAR,v1_ALICE}. Even with just one spectator plane, the momentum conservation effect can be nullified in observables like $\Delta\gamma_{112}$, as $\gamma^{\rm{OS}}_{112}$ and $\gamma^{\rm{SS}}_{112}$ are affected in the same manner~\cite{STAR-3,STAR-4}.

The background related to flow can be illustrated through the example of flowing resonances, which undergo decay into particles $\alpha$ and $\beta$~\cite{CME-9}: 
\begin{equation}
{\rm BG}_{\rm res} \propto v_{2}^{\rm res} \langle\cos(\varphi_\alpha+\varphi_\beta-2\varphi_{\rm res})\rangle.
\label{bkgrelation}
\end{equation}
As an example, consider two $\rho$ mesons traveling in opposite directions, but both within the reaction plane. In this case, $v_{2}^{\rm res} = 1$. If the decay pions are so highly boosted as to follow the exact directions of their parents, there exists no out-of-plane charge separation, yet $\Delta\gamma_{112}$ based on these four pions equals 1.
The flowing resonance picture can be generalized to a larger portion of, or even the entire, event
through the mechanisms of transverse momentum conservation (TMC)~\cite{Pratt2010,Flow_CME} 
and/or local charge conservation (LCC)~\cite{PrattSorren:2011,LCC-1}. 
Essentially,  the flow-related background in $\Delta\gamma_{112}$ can be attributed to the coupling of elliptic flow with other mechanisms, and therefore, should be effectively suppressed by reducing elliptic flow.

The event shape selection (ESS), also called event shape engineering in some literature~\cite{SergeiESE},  classifies collision events according to their shapes and maps the CME observables to a group of events with minimal flow. 
Ideally, an event shape variable based on final-state particles can accurately reflect the eccentricity of the initial overlap region. The projection to spherical events (zero eccentricity) then guarantees that $v_2$ is on average zero for all particles, whether they are primordial particles, resonances, or decay products.
However, as demonstrated in Fig.~\ref{setup}(b), the particle emission pattern may exhibit event-by-event fluctuations, resulting in minimal $v_2$ measurements despite a finite eccentricity.
In reality, it is not a trivial task to devise a feasible and effective ESS procedure.

An early ESS practice categorizes events directly using ``observed $v_2$"~\cite{LPV_STAR5}, and faces certain technical challenges in data interpretation~\cite{1st-ESS}.
A more dependable event shape variable is $q_2$~\cite{Sergeiflow1, SergeiESE,1st-ESS}, the module of the second-order flow vector, $\overrightarrow{q_2} = \frac{1}{\sqrt{N}}\bigl(\sum_{i=1}^N \cos 2\varphi_i,\sum_{i=1}^N \sin 2\varphi_i\bigr)$.
When $q_2$ is constructed from a sub-event that does not include particles of interest (POI), the correlation between this $q_{2,{\rm B}}$ and $v_{2,{\rm POI}}$ is typically weak, and even at zero $q_{2,{\rm B}}$, $v_{2,{\rm POI}}$ is still positive and sizable.
As a result, the extrapolation of $\Delta\gamma_{112}$ over a wide, unmeasured $v_{2,{\rm POI}}$ region
introduces substantial
statistical and systematic uncertainties~\cite{ALICE-2,CMS-2}. We have also demonstrated this point in Appendix~\ref{appendix2}.
To circumvent lengthy extrapolations in $v_{2}^{\rm POI}$, we adopt the approach outlined in Ref.~\cite{1st-ESS} that builds $q_2$ upon POI, $q_{2,{\rm POI}}$. 
For simplicity, we will use $q_2$ to represent $q_{2,{\rm POI}}$ throughout the paper, except in Appendix~\ref{appendix2}.
Besides the technical justification, potential longitudinal flow-plane decorrelations~\cite{torque,HydroFluc,Glasma,Dynamic,EccenDecorr} highlight the importance of maintaining POI and $q_2$ within the same rapidity region.
Furthermore, we will incorporate
information on particle pairs
and expand the ESS method into four variants.
 
In Sec.~\ref{sec:ess} we will describe 
the ESS methodology in detail.
Sec.~\ref{sec:models} briefly outlines the models to be used in the upcoming simulation studies, including a toy model~\cite{toymodel} and the event-by-event anomalous-viscous fluid dynamics (EBE-AVFD) model~\cite{AVFD1,AVFD2,AVFD4}. 
In Sec.~\ref{sec:resonancev2}, we investigate the relationship between resonance $v_2$ and both the background and, remarkably, the CME signal. 
We also perform a quantitative assessment of the efficacy of the ESS variants in mitigating the background for $\Delta \gamma_{112}$.
In Sec.~\ref{sec:conclusion} we 
provide a summary and offer an outlook.

\section{Event Shape Selection}
\label{sec:ess}

Prior studies~\cite{1st-ESS} have demonstrated that,  compared with $q_2$, $q_2^2$ renders a more reliable projection of $\Delta\gamma_{112}$ to the zero-flow limit. Based on the expansion,
 \begin{eqnarray}
q^2_2 & = & \frac{1}{N}
\biggl[\biggl(\sum^N_{i=1} \sin2\varphi_i\biggr)^2 + \biggl(\sum^N_{i=1} \cos2\varphi_i\biggr)^2 \biggr] \nonumber \\
& = & 1 + \frac{1}{N}\sum_{i\neq j} \cos[2(\varphi_i - \varphi_j)], \label{q2_definine}
\end{eqnarray}
we estimate the event average of $q_2^2$,
\begin{equation}
\langle q_2^2\rangle    \approx 1 + N v_2^2\{2\}. \label{q2_average}
\end{equation}
Here, $v_2\{2\}$ is the ensemble average of elliptic flow obtained from two-particle correlations.
Eqs.~(\ref{q2_definine}) and (\ref{q2_average}) reveal the link between the event shape variable and the elliptic flow variable, and warrants a correction to the normalization for $q_2$. Hereafter,
we redefine $q_2^2$ as
\begin{equation}
q^2_2 = \frac{\bigl(\sum^N_{i=1} \sin2\varphi_i\bigr)^2 + \bigl(\sum^N_{i=1} \cos2\varphi_i\bigr)^2 }{N(1+Nv_2^2\{2\})}.\label{eq:new_q}    
\end{equation}
Note that $N$ varies from event to event, whereas $v_2\{2\}$ is a static constant averaged over all events. The correction due to the new term in the normalization becomes prominent when $N$ is large.
 
As mentioned in Sec.~\ref{intro},
we construct $q_2$ upon POI, in hope that the $q_2$ and $v_2$ based on single particles largely reflect the initial geometry of the collision system. 
However, the anisotropy of particle pairs could be more pertinent to the background contribution in  $\gamma_{112}$ which correlates the reaction plane with a pair of particles, instead of single particles. 
For instance, to the background arising from flowing resonances, the $v_2$ and the corresponding $q_2$ of the parent resonances should be more relevant than those of the kindred decay daughters.
More generally, the background caused by LCC entails primordial particles that contribute as if they had hypothetical flowing progenitors.
Accordingly, we invoke the pair azimuthal angle, 
$\varphi^{\rm p}$,
from the momentum sum for each pair of POI, regardless of the existence of a genuine parent. The corresponding elliptic flow variable (pair $v_{2}$) and event shape variable (pair $q^2_{2}$) are expressed as
\begin{eqnarray}
v_{2,\rm{pair}} &=& \langle \cos(2\varphi^{\rm p} - 2\Psi_{\rm {RP}}) \rangle, \label{eq:pairv2}\\
q^2_{2,{\rm pair}} &=& \frac{\bigl(\sum^{N_{\rm pair}}_{i=1} \sin2\varphi_i^{\rm p}\bigr)^2 + \bigl(\sum^{N_{\rm pair}}_{i=1} \cos2\varphi_i^{\rm p}\bigr)^2 }{N_{\rm pair}(1+N_{\rm pair}v_{2,{\rm pair}}^2\{2\})}, \label{eq:pairq2}
\end{eqnarray}
respectively.    
In Eq.~(\ref{eq:pairq2}), the correction term in the normalization is crucial, given that $N_{\rm pair}$ is basically on the order of $N^2/2$.

Our ESS analysis closely follows the procedure described in Ref.~\cite{ESS-1}. Firstly, we group events according to the event shape variable and compute both $\Delta\gamma_{112}$ and the elliptic flow variable for each group. 
Next, we study the dependence of $\Delta\gamma_{112}$ on the elliptic flow variable and project it to the zero-flow limit.
Now that both the event shape variable and the elliptic flow variable can be calculated either from single particles or from particle pairs, we will discuss the following four  combinations. 

\begin{tikzpicture}
\begin{scope}%[every node/.style={squarenode,minimum size=.75cm,draw}]
    \node (o) at (0,0) {};
    \node (a) at (2,0) {single $q^2_2$};
    \node (b) at (5.5,0) {single $v_2$};
    \node (c) at (2,-1) {pair $q^2_2$};
    \node (d) at (5.5,-1) {pair $v_2$};
\end{scope}
\draw (a)--(b) (c)--(d);
\draw [dashed] (a)--(d)  (c)--(b);
\end{tikzpicture}

\noindent
Since the unmixed recipes (with single $q_2^2$ and single $v_2$ or with pair $q_2^2$ and pair $v_2$) rely on the azimuthal angles of the same POI or the same pairs, there could be intrinsic correlations between the event shape variable and the elliptic flow variable, as indicated by 
Eqs.~(\ref{q2_definine}) and (\ref{q2_average}). Therefore, the projection to the zero-flow limit may be prone to bias, causing a residual background. Conversely, the mixed recipes (with single $q_2^2$ and pair $v_2$ or with pair $q_2^2$ and single $v_2$)
enhance the mutual independence between the event shape variable and the elliptic flow variable, and could improve the extrapolation.
The sensitivity of all four approaches to background subtraction will be tested in the forthcoming sections.

\section{Model Descriptions} \label{sec:models}
The simple toy model~\cite{toymodel} without the CME aims to  investigate the background mechanism due to flowing resonances.
In each event, we generate 33 pairs of $\pi^\pm$ that come from $\rho$ meson decays, the multiplicity of which roughly matches the STAR measurements  within the rapidity range of $|y|<1$ in 30--40\% Au+Au collisions at $\sqrt{s_{NN}}$ = 200 GeV~\cite{mult_200,percent_17}.  
The inputs of spectrum and elliptic flow for $\rho$ mesons follow the same procedure as adopted in Ref.~\cite{toymodel}.
%The $\rho$-meson spectrum obeys $ \frac{dN_{\rho}}{dm^{2}_{T}} \propto \frac{e^{-(m_{T}-m_{\rho})/T}}{T(m_{\rho}+T)}$, with $T = 317$~MeV to reproduce the observed $\langle p_{T} \rangle$ of 830 MeV/$c$~\cite{percent_17}. 
The decay process, $\rho \rightarrow \pi^{+} + \pi^{-}$, is implemented with PYTHIA6~\cite{pythia6}. In this study, we only select pions within $0.2 < p_T < 2$ GeV/$c$. 

%%%%%%%%%%%%%%%%%
%\subsection{AVFD Model}
The EBE-AVFD model~\cite{AVFD1,AVFD2,AVFD4}  incorporates the dynamical CME transport for light quarks,
and effectively handles the background mechanisms like flowing resonances, TMC, and LCC.
The initial conditions for entropy density ($s$) profiles and for electromagnetic field follow the event-by-event nucleon configuration from the Monte Carlo Glauber simulations~\cite{glauber}.
The input of  chirality charge density ($n_5$) takes the form of $n_5/s$.  
The CME transport is governed by anomalous hydrodynamic equations as a linear perturbation on top of the medium flow that is managed by the VISH2+1 simulation package~\cite{AVFD3}. In the freeze-out process, the LCC effect is specifically tuned to match experimental data. 
After being produced from the freeze-out hypersurface, all hadrons undergo hadron cascades using the UrQMD simulations~\cite{Bleicher:1999xi}, which account for various resonance decay processes and automatically factor in their contributions to the charge-dependent correlations.
We focus on the 30--40\% centrality range of Au+Au collisions at $\sqrt{s_{NN}} = 200$ GeV, using the same settings and input parameters as adopted in Ref.~\cite{AVFD5}. The numbers of events are $9.6 \times10^7$, $5.9 \times 10^7$, and $7.7 \times 10^7$ for the cases of $n_5/s=0$, 0.1, and 0.2, respectively. For simplicity, the analysis uses the true reaction plane.
The POI are $\pi^\pm$ within  $|y|<1$ and  $0.2 < p_T < 2$ GeV/$c$.

\section{simulation results} \label{sec:resonancev2}
%In this section, we focus on understanding the root cause of the non-CME background in $\Delta\gamma_{112}$, which is represented by resonance $v_2$, and examine its influence on ESS variables. 

\begin{figure}[bthp]
\centering
\includegraphics[scale=0.4]{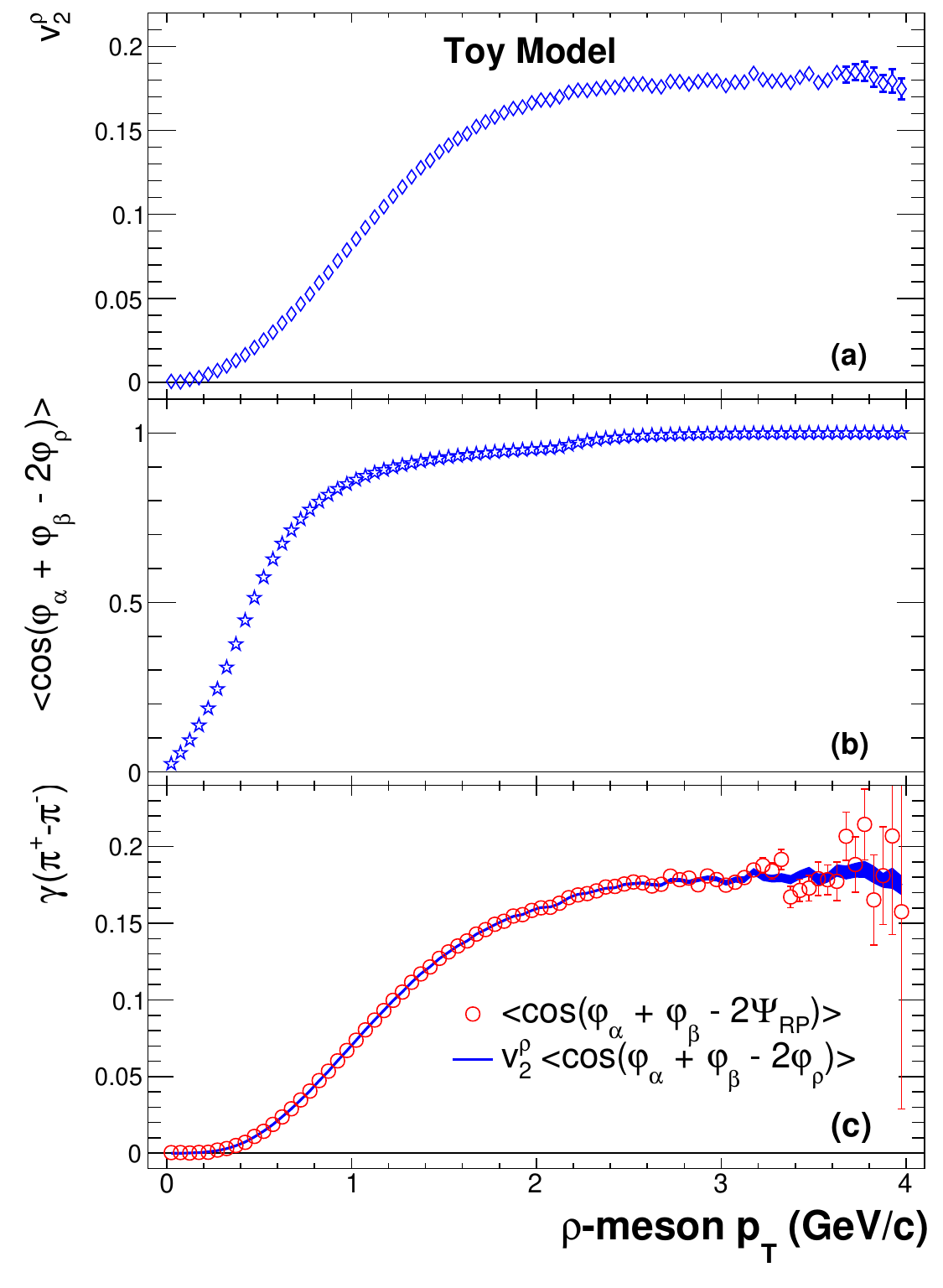}
\caption{Toy model simulations of (a) $\rho$-meson $v_2$, (b) the correlation between a $\rho$ meson and its daughter pions, $\langle \cos(\varphi_\alpha +\varphi_\beta - 2\varphi_\rho) \rangle$, and (c) the $\gamma_{112}$ correlator between two sibling pions, as a function of $\rho$-meson $p_T$. For comparison, the product, $v_2^\rho\langle \cos(\varphi_\alpha +\varphi_\beta - 2\varphi_\rho) \rangle$, is shown with a shaded band.}
\label{gamma_rho_toy}
\end{figure}

Figure~\ref{gamma_rho_toy}(a) shows the input of $\rho$-meson $v_2$ for the toy model, and Fig.~\ref{gamma_rho_toy}(b) presents the  correlation between a $\rho$ meson and its daughter pions, $\langle \cos(\varphi_\alpha +\varphi_\beta - 2\varphi_\rho) \rangle$, which is solely due to decay kinematics. As $\rho$-meson $p_T$ increases, 
the two daughters display enhanced collimation towards the direction of the parent's motion.
Figure~\ref{gamma_rho_toy}(c) shows the resultant $\gamma_{112}$ correlator between two sibling pions,
which agrees very well with the product,  
$v_2^\rho\langle \cos(\varphi_\alpha +\varphi_\beta - 2\varphi_\rho) \rangle$.
This agreement corroborates the relation stated in Eq.~(\ref{bkgrelation}) and  
confirms the background mechanism due to the decay of flowing resonances.

\begin{figure}[bthp]
\centering
\includegraphics[scale=0.4]{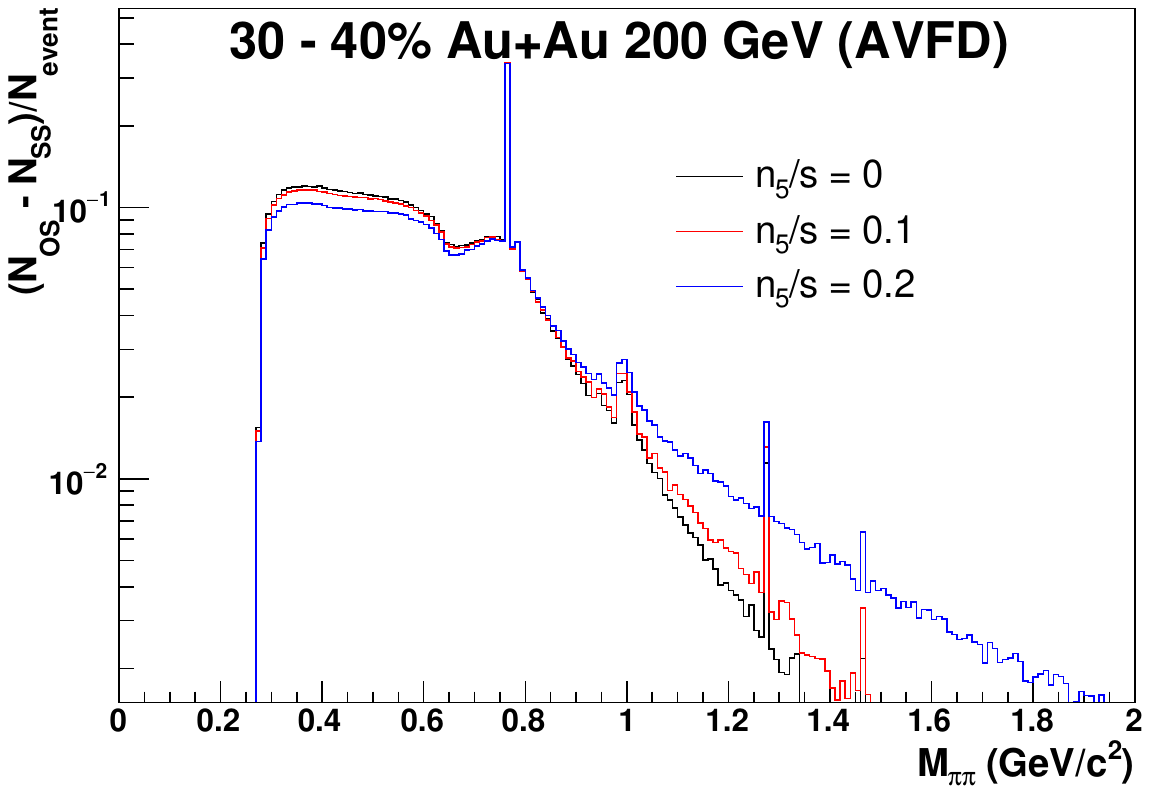}
\caption{EBE-AVFD calculations of the  per-event-normalized excess of opposite-sign $\pi$-$\pi$ pairs over same-sign ones vs   invariant mass in 30--40\% Au+Au collisions at 200 GeV. In the low-mass region, the results from top to bottom correspond to the cases of $n_5/s =$ 0, 0.1, and 0.2, respectively.
The ordering is inverted in the high-mass region. 
The sharp peaks resemble resonances such as $\rho$ meson.}
\label{InvariantMass_avfd}
\end{figure}

Given the consistently positive nature of $\langle \cos(\varphi_\alpha +\varphi_\beta - 2\varphi_\rho) \rangle$ without any sign change, 
it is tempting to directly use resonance $v_2$ as the elliptic flow variable for background control.
However, we realize that the experimental manifestation of resonances, the excess of OS over SS particle pairs, is affected by the CME. 
Figure~\ref{InvariantMass_avfd} delineates the EBE-AVFD simulations of the per-event-normalized invariant mass distribution for the excess of OS  over SS pion pairs in 30--40\% Au+Au collisions at 200 GeV. 
The invariant mass of two pions is  intimately linked to the opening angle between 
them. The presence of the CME, characterized by finite $n_5/s$ values, tends to induce an opposing motion of $\pi^+$ and $\pi^-$, enhancing the observed resonance yield at higher invariant mass, and meanwhile depleting the lower-mass region.

\begin{figure}[bthp]
\centering
\includegraphics[scale=0.44]{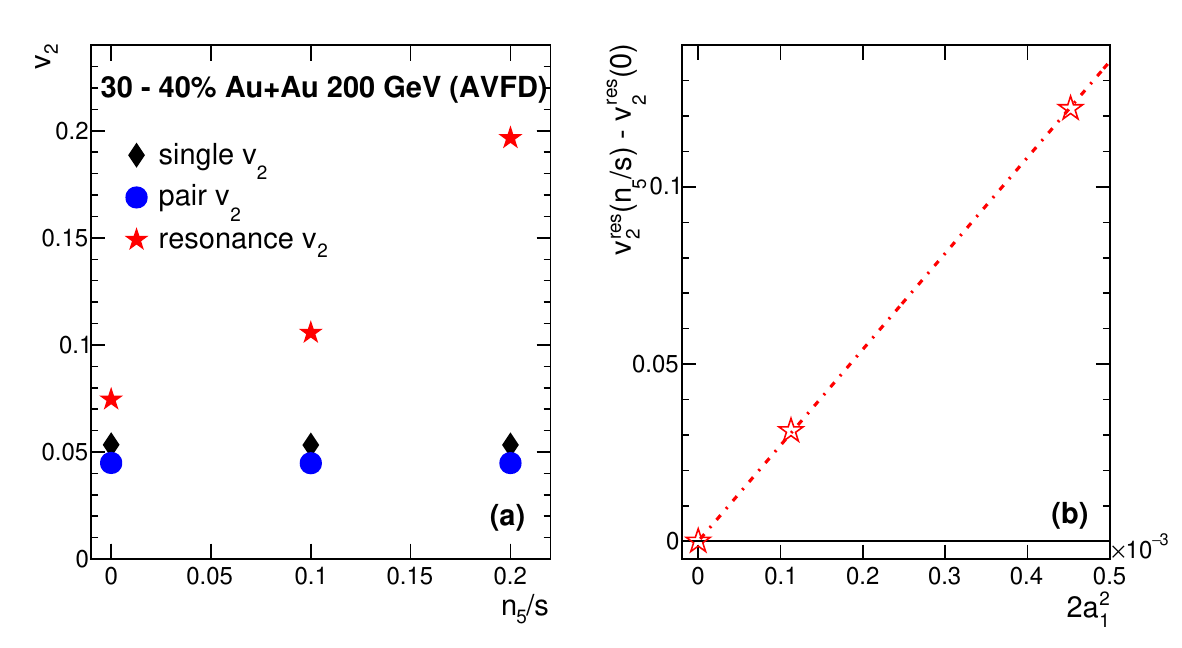}
\caption{(a) EBE-AVFD simulations of single $v_2$, pair $v_2$, and resonance $v_2$ as a function of $n_5/s$ in 30--40\% Au+Au collisions at 200 GeV. (b) Background subtracted resonance $v_2$ vs $2a_1^2$. The linear fit passing through $(0, 0)$ demonstrates that resonance $v_2$ contains the CME signal proportional to $a_1^2$.}
\label{v2_avfd}
\end{figure}

The CME  not only impacts the invariant mass spectrum of observed resonances, but also modifies the corresponding $v_2$ observable, 
\begin{equation}
v_2^{\rm res} \equiv \frac{N_{\rm{OS}} \langle\cos2\Delta\varphi^{\rm{OS}}\rangle - N_{\rm{SS}}\langle\cos2\Delta\varphi^{\rm{SS} }\rangle}{N_{\rm{OS}} - N_{\rm{SS}}}, \label{eq:res_v2}
\end{equation}where $\Delta\varphi^{\rm{OS(SS)}}$ denotes the azimuthal angle of an OS(SS) particle pair with respect to $\Psi_{\rm RP}$. Figure~\ref{v2_avfd}(a) shows the EBE-AVFD calculations of single $v_2$, pair $v_2$, and resonance $v_2$ as a function of $n_5/s$. The former two  essentially remain the same, whereas resonance $v_2$ linearly increases with the strength of the CME.
Here, we take the entire invariant mass range  without selecting any specific resonance.
Figure~\ref{v2_avfd}(b) presents the background subtracted resonance $v_2$ vs $2a_1^2$.
We apply a linear fit passing through $(0, 0)$ to  demonstrate that resonance $v_2$ contains the CME signal proportional to $a_1^2$.

%The resonance $v_2^{res}$ is subtracted from the excess pair angle distribution $(\varphi^{\rm{p}}-\Psi_{RP})$ as the second-order Fourier coefficient. 
%The combinatorial background is nullified by the subtraction of SS pairs from OS pairs, following the standard experimental approaches in resonance reconstruction.

\begin{figure}[tbph]
\begin{minipage}[c]{0.235\textwidth}
\includegraphics[width=\textwidth]{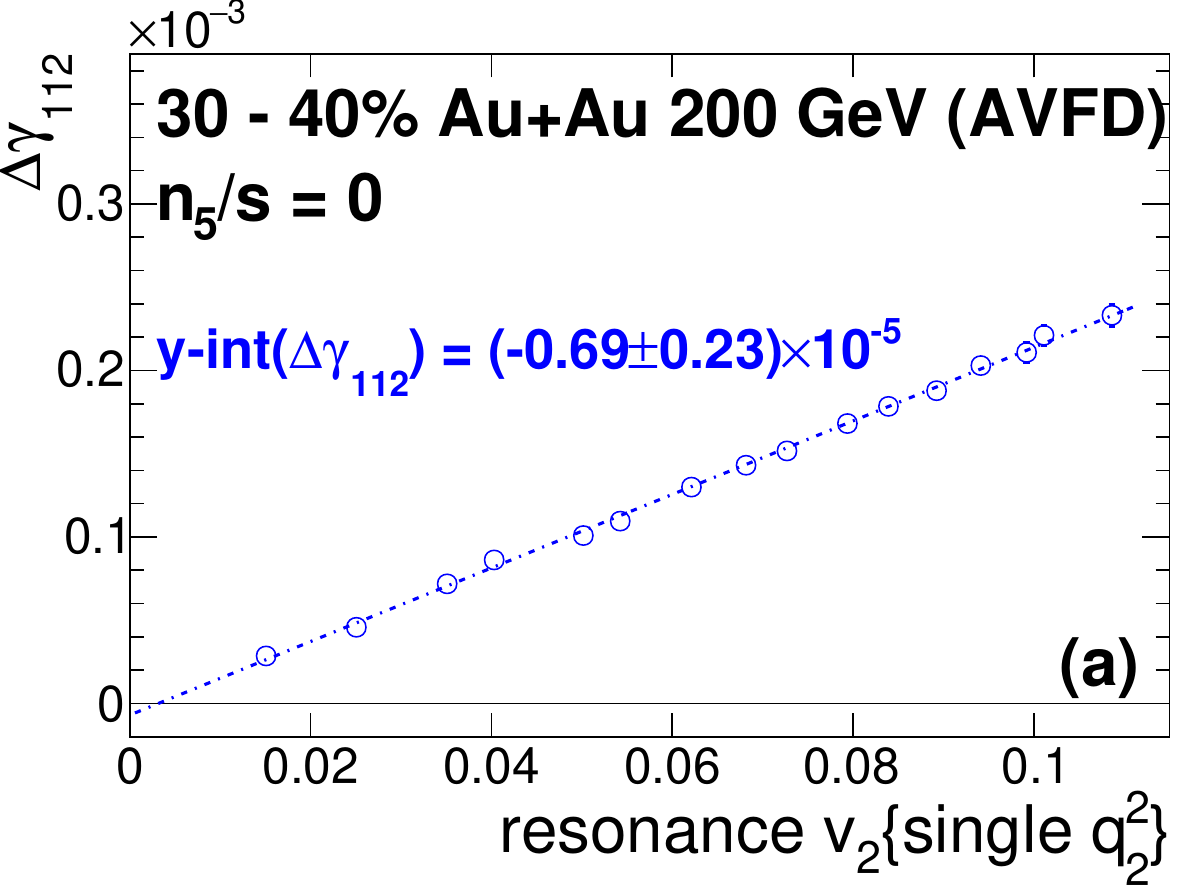}
\end{minipage}
\begin{minipage}[c]{0.235\textwidth}
\includegraphics[width=\textwidth]{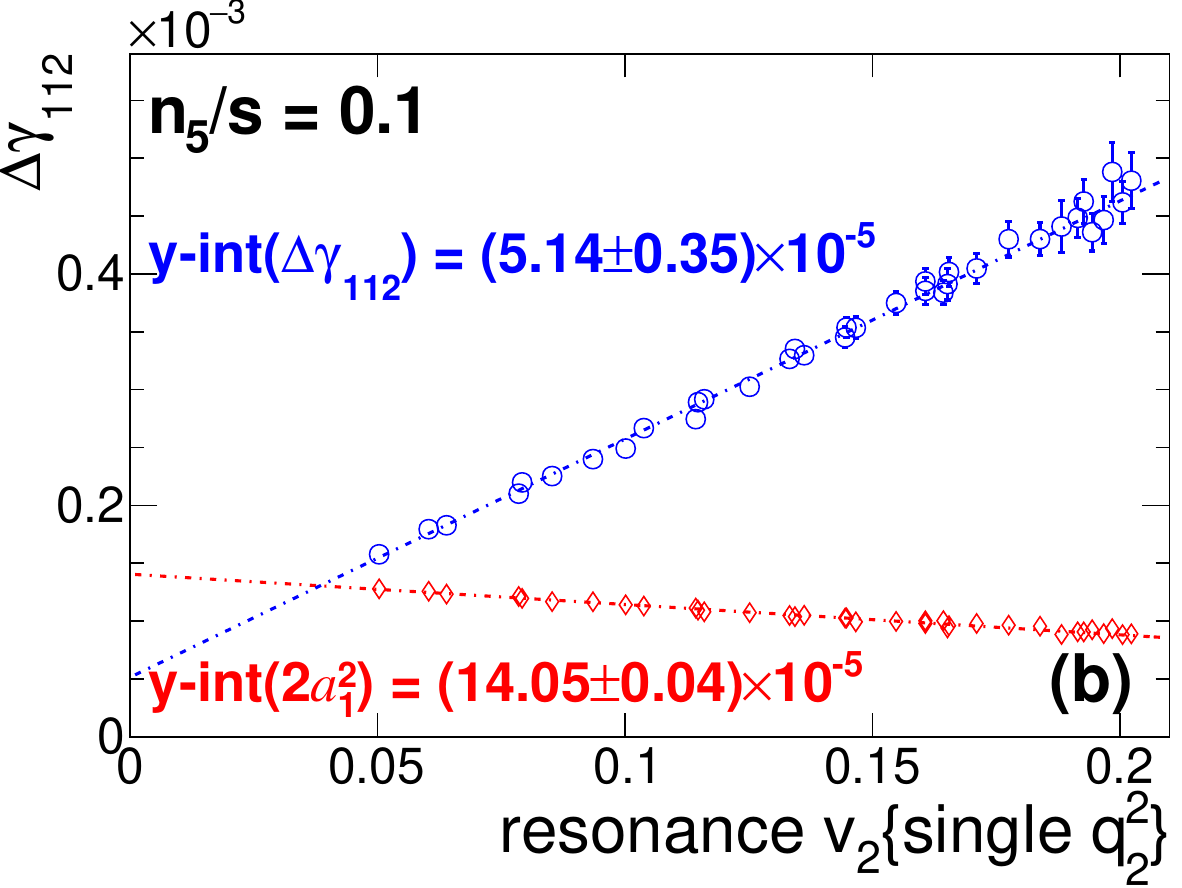}
\end{minipage} 
\caption{EBE-AVFD simulations of  $\Delta\gamma_{112}$ as a function of resonance $v_2$ categorized by single $q_{2}^2$ in the  scenarios of $n_5/s=0$ (a) and 0.1 (b) in 30--40\% Au+Au collisions at 200 GeV. The particles of interest are  charged pions. For comparison, $2a_1^2$ is also added  in panel (b). The  $y$-intercepts obtained with linear fits demonstrate the over-subtraction of the elliptic-flow-related background.  
}
\label{AVFD_4}
\end{figure}

As derived analytically in Appendix~\ref{appendix1} with certain simplifications, resonance $v_2$ is another CME  observable,
\begin{eqnarray}
v_2^{\rm res} 
&\approx& v_2^{\pi} +\frac{N_{\rm{OS}}+N_{\rm{SS}}}{2(N_{\rm{OS}}-N_{\rm{SS}})}(\Delta\gamma_{112}-v_2^\pi\Delta\delta) \nonumber \\
& & +\frac{1}{2}(\gamma^{\rm OS} +\gamma^{\rm SS}) - \frac{v_2^\pi}{2}(\delta^{\rm OS}+\delta^{\rm SS}),
\label{resonancev2_2}
\end{eqnarray}
where  $\delta \equiv \langle \cos(\varphi_\alpha-\varphi_\beta) \rangle$ is the two-particle correlation.
Since  $(N_{\rm{OS}}+N_{\rm SS})/(N_{\rm{OS}}-N_{\rm{SS}})$ is typically on the order of a few hundred,   the second term in Eq.~(\ref{resonancev2_2}) is comparable to  $v_2^{\pi}$. This makes resonance $v_2$ unsuitable as an elliptic flow variable in the ESS method. Figure~\ref{AVFD_4} shows the EBE-AVFD simulations of $\pi$-$\pi$ $\Delta\gamma_{112}$ as a function of  $v_2^{\rm res}$ categorized by single $q_{2}^2$ in the  scenarios of $n_5/s=0$ (a) and 0.1 (b).
The $y$-intercepts obtained with linear fits reveal the over-subtraction of background,  especially pronounced when the true CME signal, $2a_1^2$, is finite.

Similarly, pair $v_2$ also contains the CME signal:
\begin{eqnarray}
v_{2,{\rm pair}} &=&\frac{N_{\rm{OS}} \langle\cos2\Delta\varphi^{\rm{OS}}\rangle + N_{\rm{SS}}\langle\cos2\Delta\varphi^{\rm{SS} }\rangle}{N_{\rm{OS}} + N_{\rm{SS}}}\\
&\approx& v_2^{\pi} +\frac{N_{\rm{OS}}-N_{\rm{SS}}}{2(N_{\rm{OS}}+N_{\rm{SS}})}(\Delta\gamma_{112}-v_2^\pi\Delta\delta) \nonumber\\
& & +\frac{1}{2}(\gamma^{\rm OS} +\gamma^{\rm SS}) - \frac{v_2^\pi}{2}(\delta^{\rm OS}+\delta^{\rm SS}).
\label{resonancev2_15}
\end{eqnarray}
Considering the small values of $(N_{\rm OS}-N_{\rm{SS}})/(N_{\rm{OS}}+N_{\rm{SS}})$ and later terms, 
pair $v_2$ is evidently dominated by $v_2^\pi$, leading to the observed constancy of pair $v_2$ in Fig.~\ref{v2_avfd}(a).  However, when using pair $v_2$ as the  elliptic flow variable for background control, the second term in Eq.~(\ref{resonancev2_15}) becomes nonnegligible, because the over-subtraction of background as demonstrated in Fig.~\ref{AVFD_4} is still present, although to a lesser extent.

Previous works using EBE-AVFD~\cite{ESS-1} have shown that 
constructing both $v_2$ and $q_2$ from the same group of particles can introduce inherent self-correlations.
Consequently, the unmixed ESS variants 
(with single $q_2^2$ and single $v_2$ or with pair $q_2^2$ and pair $v_2$) tend to under-subtract the background.
To mitigate such residual backgrounds, we have introduced in Sec.~\ref{sec:ess} mixed combinations (with pair $q_2^2$ and single $v_2$ or with single $q_2^2$ and pair $v_2$).  However, as mentioned in the preceding paragraph,  pair $v_2$   also encompasses the CME signal, which could cause an over-subtraction of background in  $\Delta\gamma_{112}$. 
Therefore,  the ESS recipe using  single $v_2$ as the elliptic flow variable and  pair $q_2^2$ as the event shape variable emerges as the optimal solution.

%%%%%%%%%
%\section{$\Delta\gamma$ measurement} \label{sec:dg112}

%One of the ESS recipe using single $q_2^2$ binning and single $v_2$ has been reported in Ref.~\cite{ESS-1} that demonstrated the ability to suppress all flow-related background in AMPT events. 
%In this section, we extend our investigation to the four ESS recipes in different realistic simulations.
%We conducted simulations with AMPT and find that all four ESS work well in terms of successfully rebuilding the pure background case for $\Delta \gamma_{112}$. The plots are not included for simplicity. 
%We focus the realistic case with true CME signal that are provided by the AVFD model.
\begin{figure}[tbph]
\begin{minipage}[c]{0.235\textwidth}
\includegraphics[width=\textwidth]{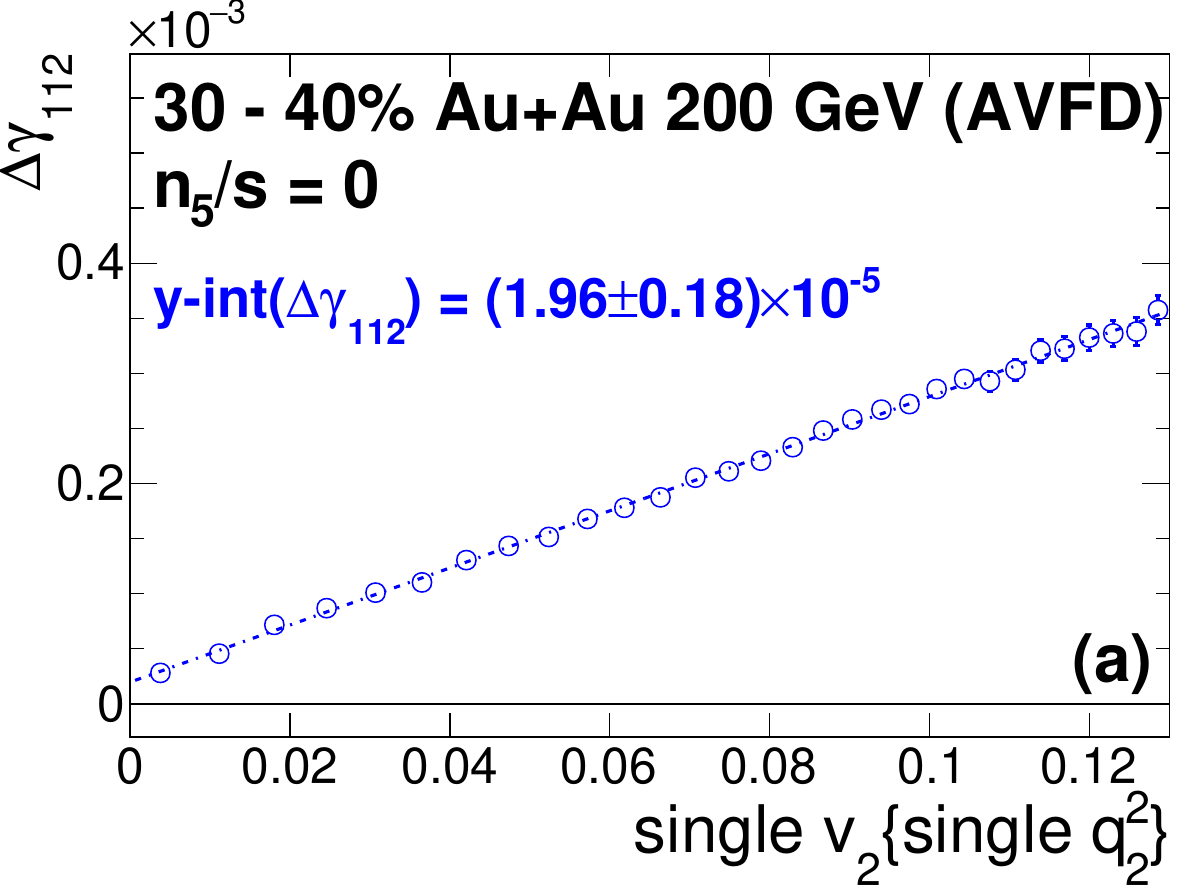}

\end{minipage}
\begin{minipage}[c]{0.235\textwidth}
\includegraphics[width=\textwidth]{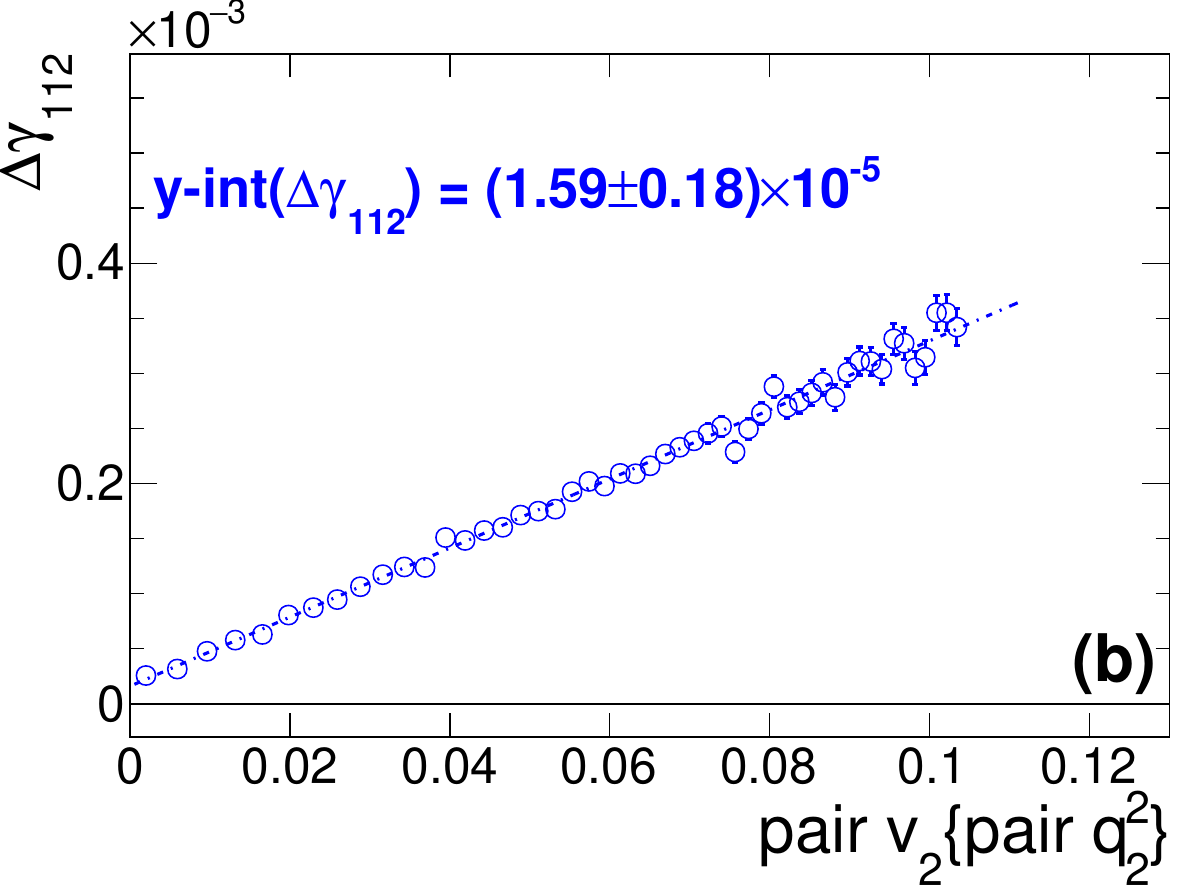}
\end{minipage} 
\begin{minipage}[c]{0.235\textwidth}
\includegraphics[width=\textwidth]{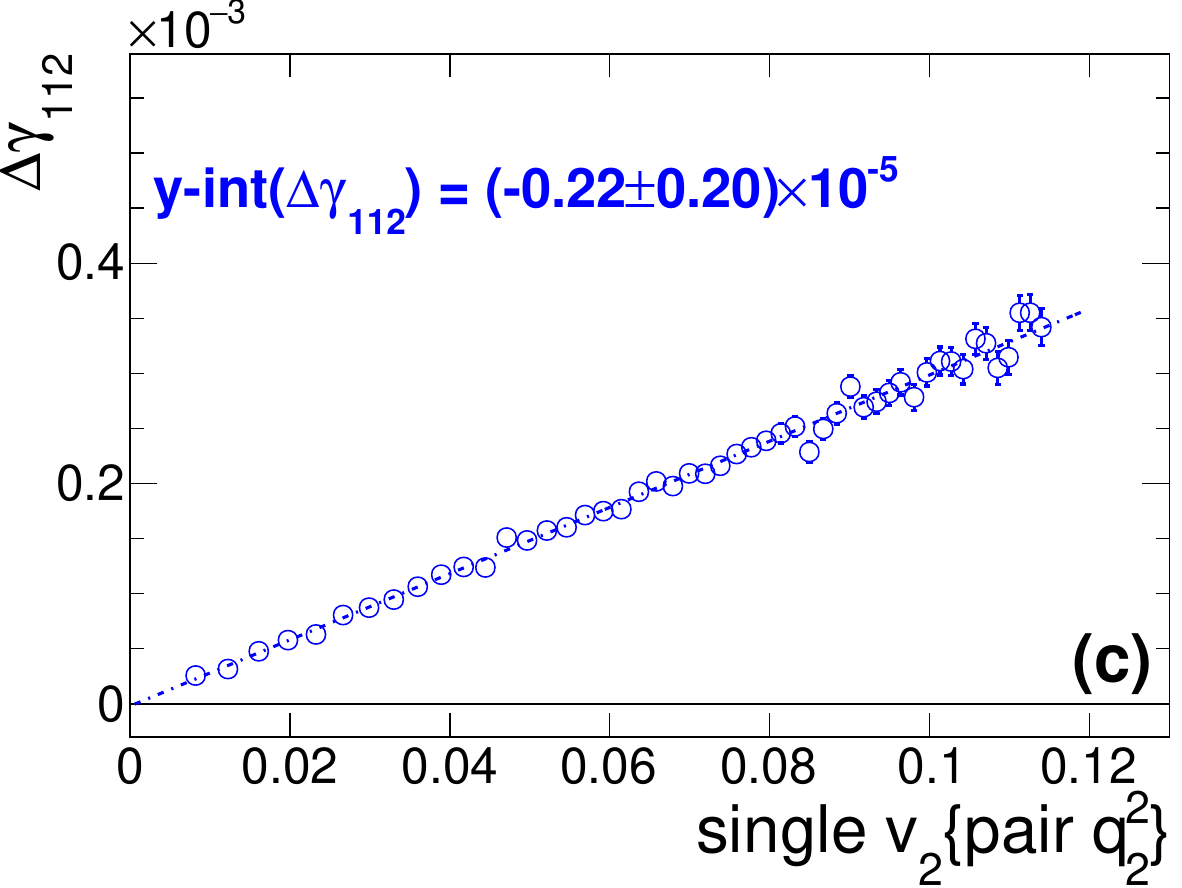}

\end{minipage}
\begin{minipage}[c]{0.235\textwidth}
\includegraphics[width=\textwidth]{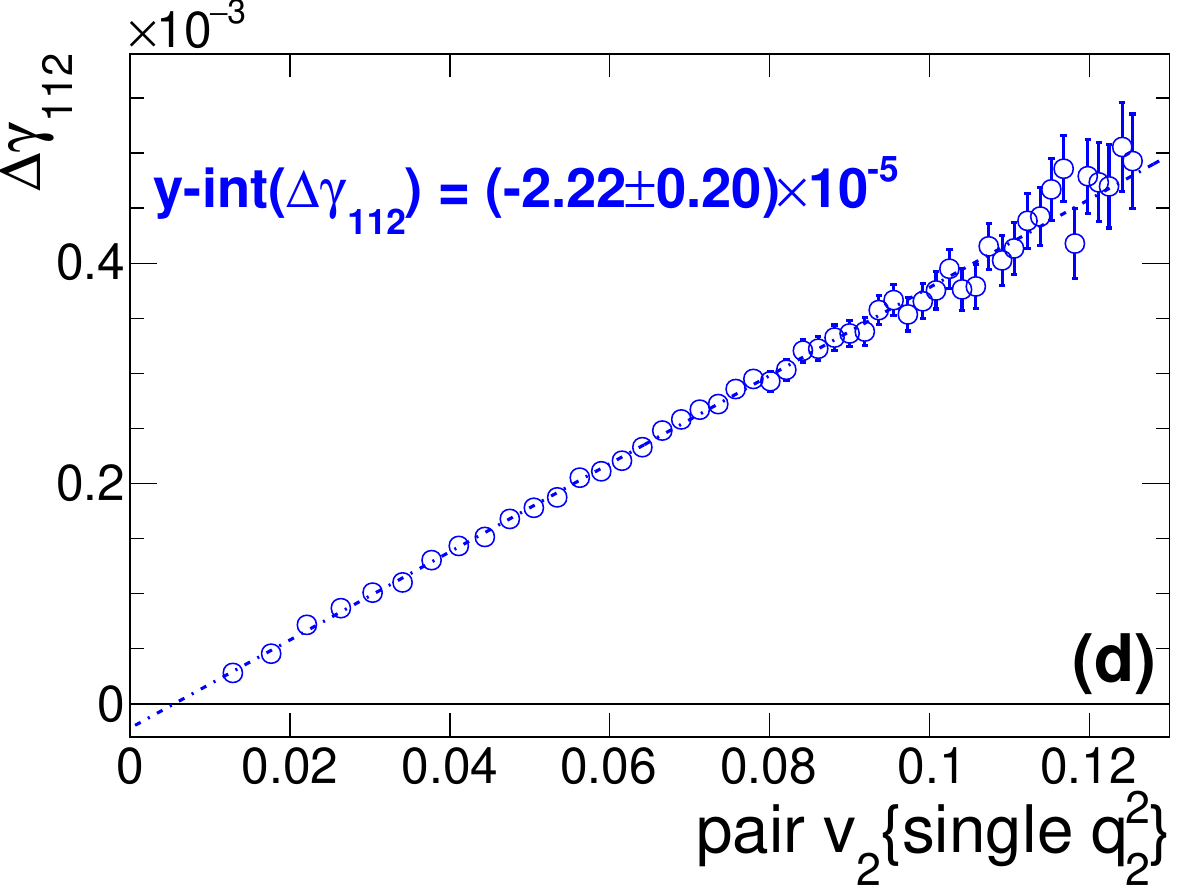}
\end{minipage} 
\caption{EBE-AVFD simulations of $\Delta\gamma_{112}$ as a function of single or pair $v_2$ categorized by single or pair $q_2^2$ in the pure background scenario ($n_5/s=0$) in 30--40\% Au+Au collisions at 200 GeV. The particles of interest are  charged pions. The $y$-intercepts obtained with linear fits reflect the efficacy of these ESS recipes in reducing the elliptic-flow-related background.
}
\label{AVFD_1}
\end{figure}

Figure~\ref{AVFD_1} shows the EBE-AVFD simulations of the $\pi$-$\pi$ $\Delta\gamma_{112}$ correlation 
as a function of $v_2$ categorized by $q_2^2$ in the pure background scenario ($n_5/s=0$) in 30--40\% Au+Au collisions at 200 GeV.  Since both $v_2$ and $q_2^2$ can be computed from either single particles or particle pairs, we present each of the four cases in a separate panel.  
%The ensemble average of $\Delta\gamma_{112}$ is $(15.68\pm 0.09)\times10^{-5}$, larger than the corresponding values in both the Toy Model and AMPT. 
The $y$-intercepts obtained from single $v_2$\{single $q_2^2$\} and  pair $v_2$\{pair $q_2^2$\} reflect an approximate eight-fold suppression of background, but they remain finite, indicative of residual backgrounds.
The ESS result using single $v_2$\{pair $q_2^2$\} is consistent with zero, whereas the one using pair $v_2$\{single $q_2^2$\} appears to over-subtract the background, as expected.

\begin{figure}[tbhp]
\begin{minipage}[c]{0.235\textwidth}
\includegraphics[width=\textwidth]{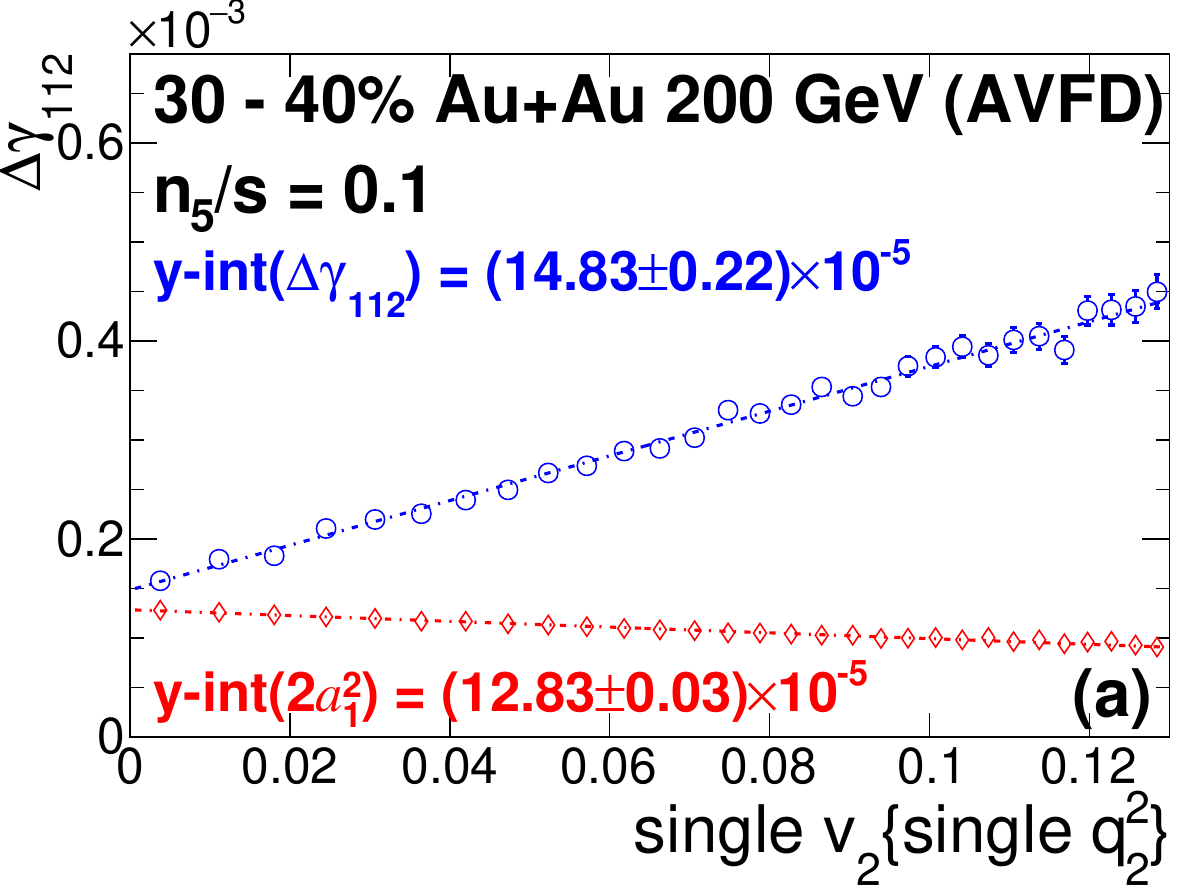}

\end{minipage}
\begin{minipage}[c]{0.235\textwidth}
\includegraphics[width=\textwidth]{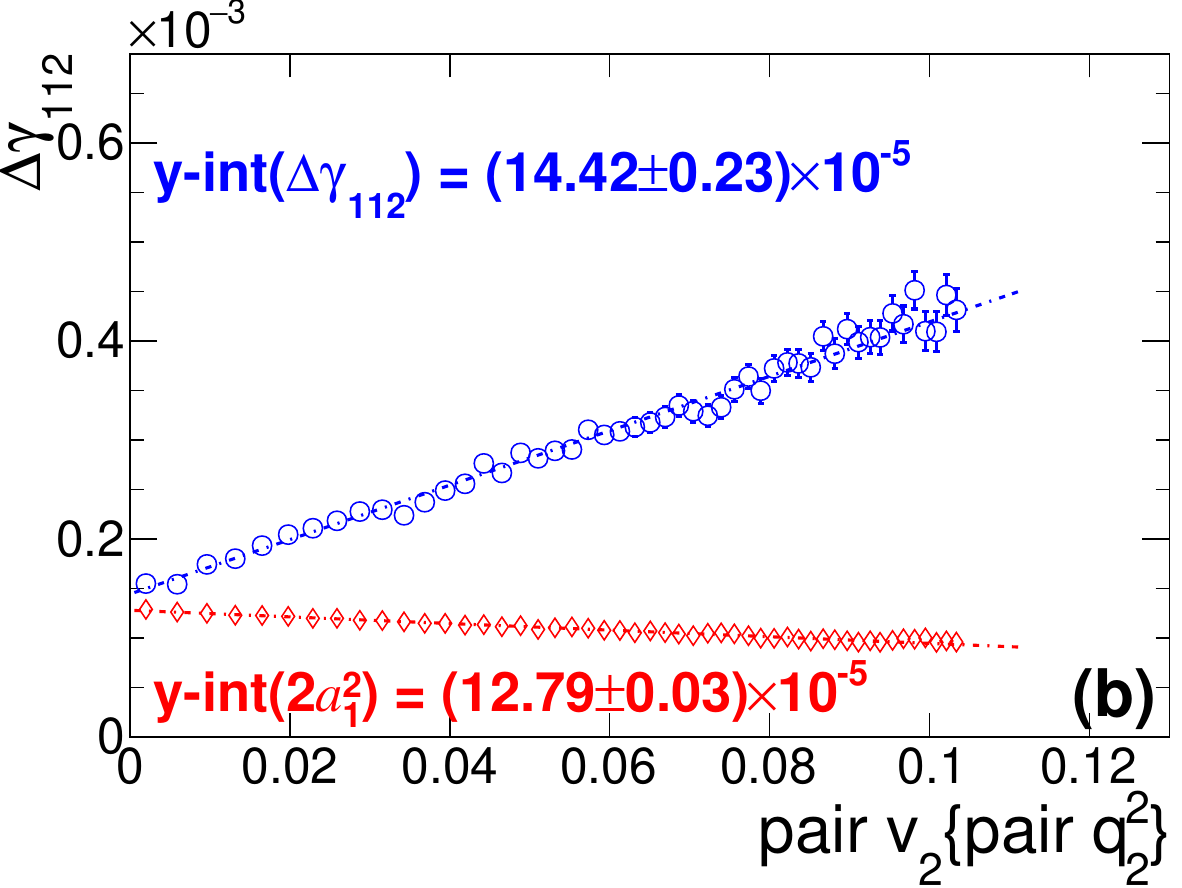}
\end{minipage} 
\begin{minipage}[c]{0.235\textwidth}
\includegraphics[width=\textwidth]{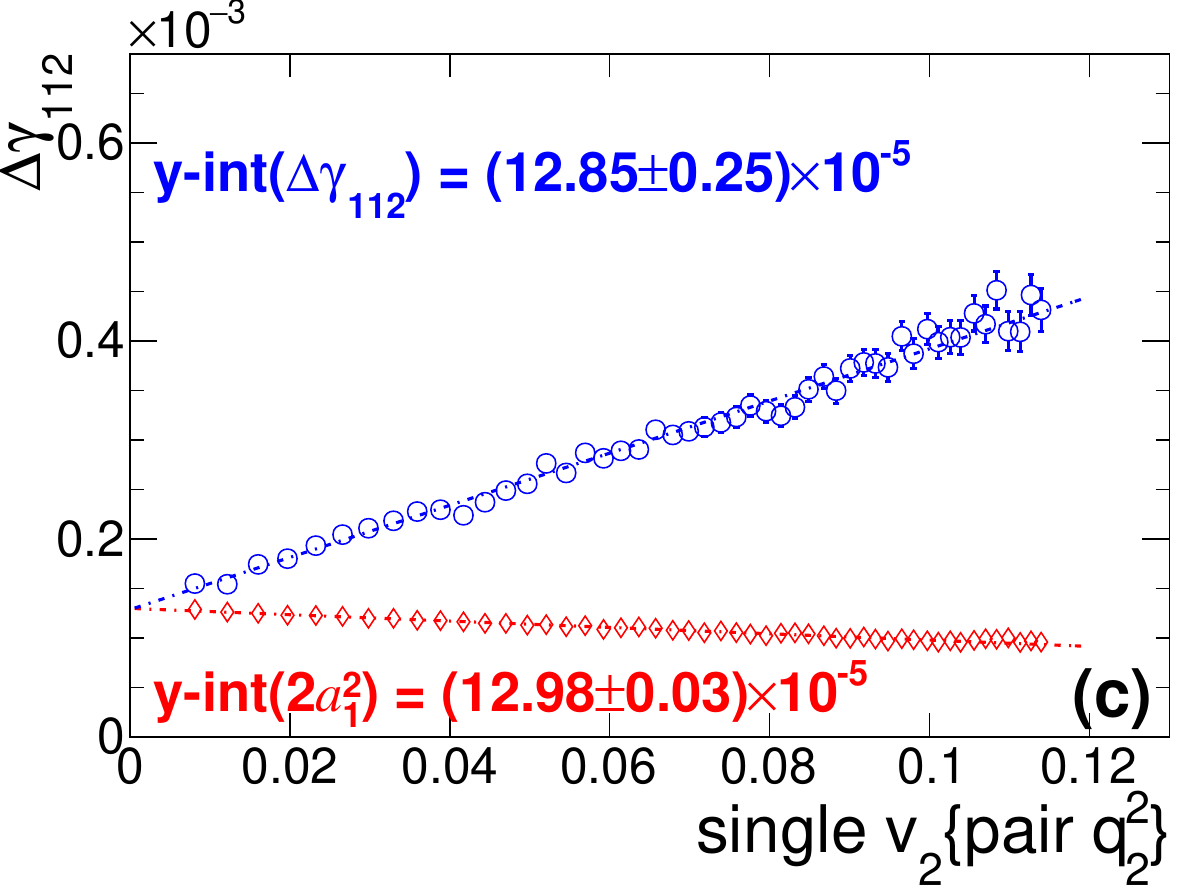}

\end{minipage}
\begin{minipage}[c]{0.235\textwidth}
\includegraphics[width=\textwidth]{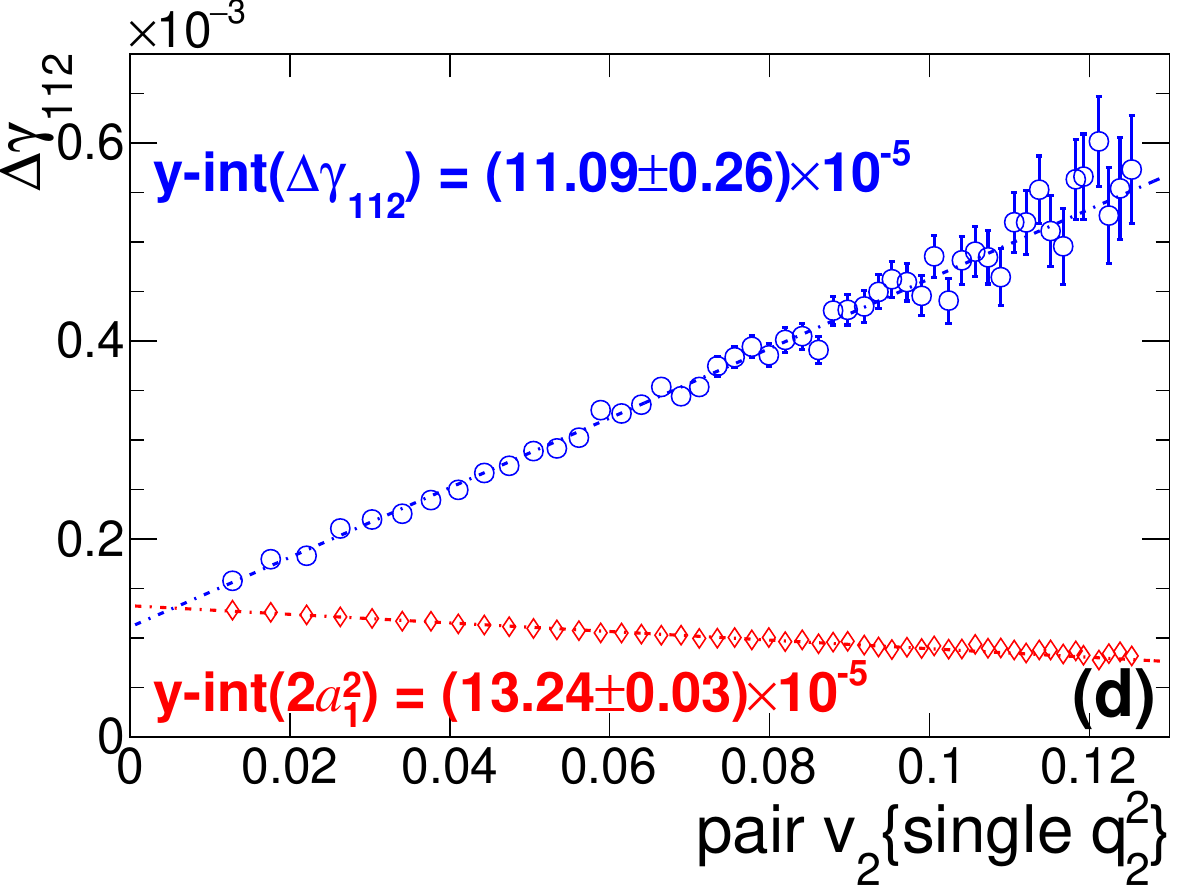}
\end{minipage} 
\caption{EBE-AVFD simulations of $\Delta\gamma_{112}$ (blue circle) and $2a_{1}^2$ (red diamond) as a function of single or pair $v_2$ categorized by single or pair $q_2^2$ in the moderate CME scenario ($n_5/s=0.1$) in 30--40\% Au+Au collisions at 200 GeV. The particles of interest are charged pions. The $y$-intercepts obtained with linear fits reflect the efficacy of these ESS recipes in revealing the true CME signal.
}
\label{AVFD_2}
\end{figure}

\begin{figure}[bthp]
\begin{minipage}[c]{0.235\textwidth}
\includegraphics[width=\textwidth]{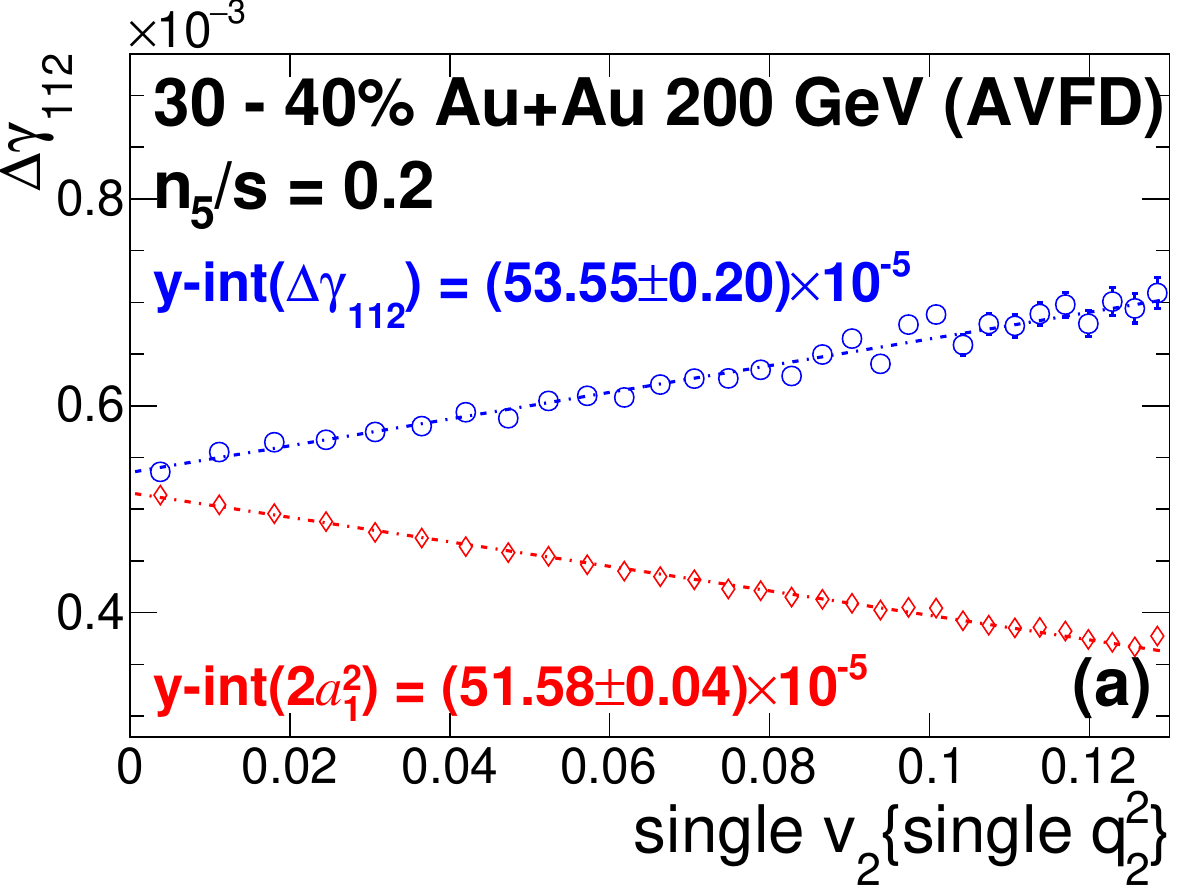}

\end{minipage}
\begin{minipage}[c]{0.235\textwidth}
\includegraphics[width=\textwidth]{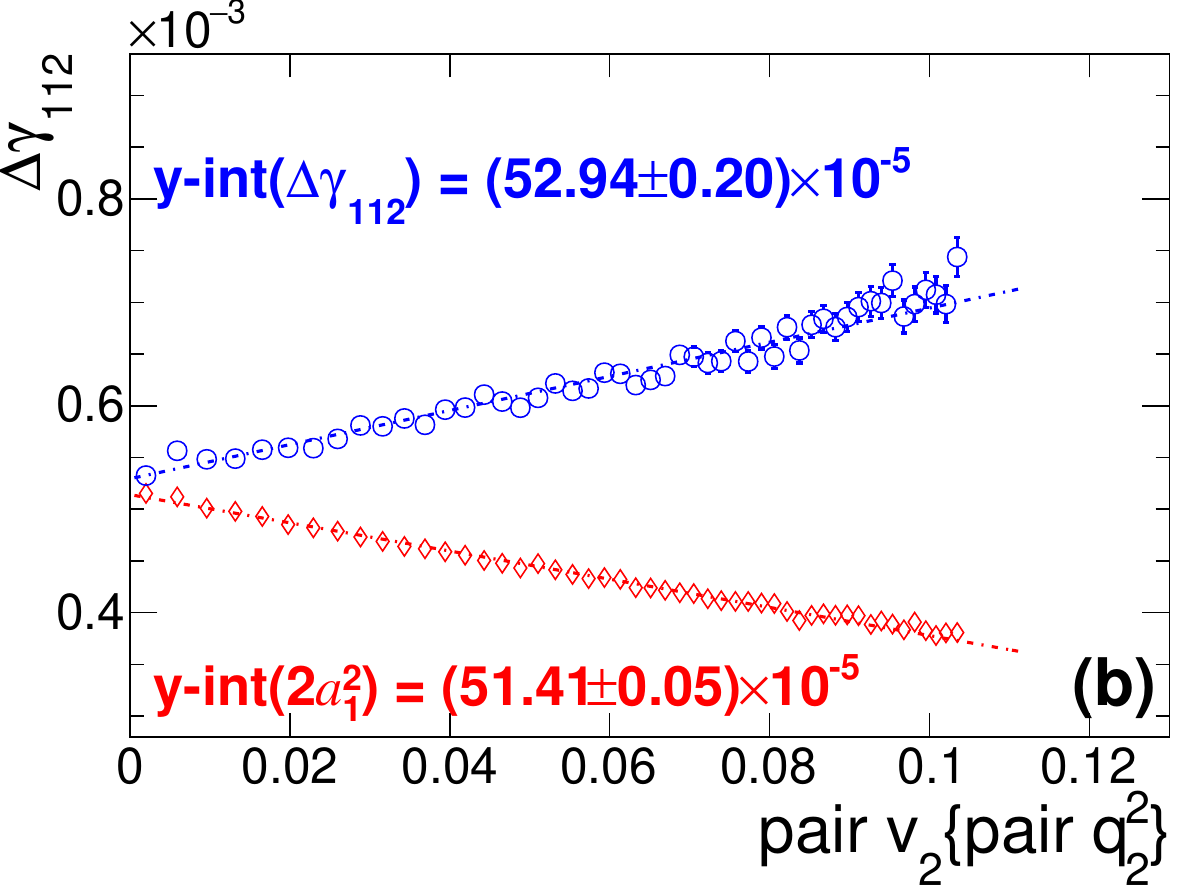}
\end{minipage} 
\begin{minipage}[c]{0.235\textwidth}
\includegraphics[width=\textwidth]{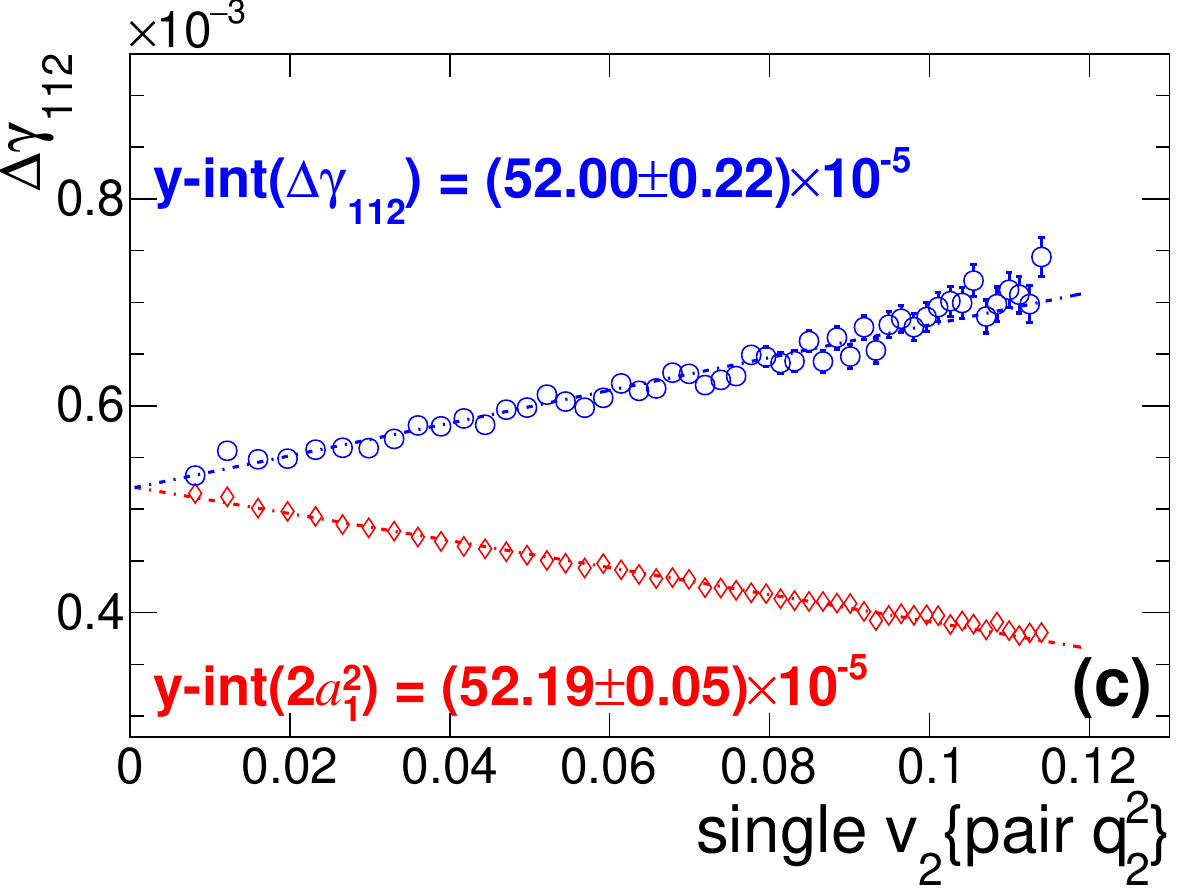}

\end{minipage}
\begin{minipage}[c]{0.235\textwidth}
\includegraphics[width=\textwidth]{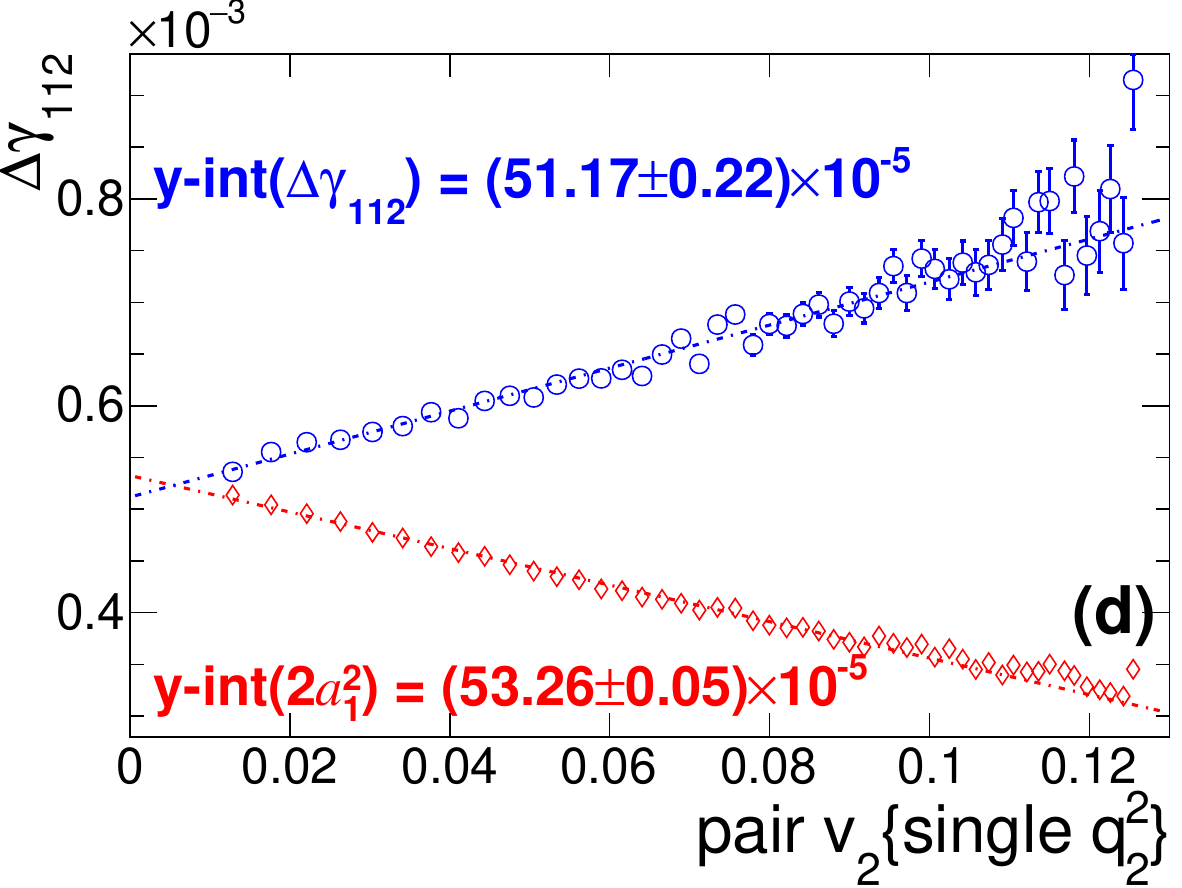}
\end{minipage} 
\caption{The same as in Fig.~\ref{AVFD_2}, but for the strong CME scenario ($n_5/s=0.2$).
}
\label{AVFD_3}
\end{figure}

Figures~\ref{AVFD_2} and \ref{AVFD_3} depict scenarios that are analogous to Fig.~\ref{AVFD_1}, but for moderate and strong CME signals with $n_5/s=0.1$ and 0.2, respectively.
For comparison, we also project the true CME signal, $2a_1^2$, to zero $v_2$.
In both scenarios, the $y$-intercepts of $\Delta\gamma_{112}$ obtained from single $v_2$\{single $q_2^2$\} and  pair $v_2$\{pair $q_2^2$\}
are significantly above the actual signal, while that using pair $v_2$\{single $q_2^2$\}  overestimates the background. The analyses using single $v_2$\{pair $q_2^2$\} seem to recover the true signal within the uncertainties.
Therefore, the ESS outcomes are most reliable when we use pair $q_2^2$
for event shape selection and single-particle $v_2$ for background control.

As portrayed in Figs.~\ref{AVFD_2} and \ref{AVFD_3}, $a_1^2$ itself increases during the ESS extrapolation to the zero-flow limit.
To restore  the ensemble average of the CME signal, we need to scale the $y$-intercept of $\Delta\gamma_{112}$ with an appropriate  factor. Recent research, based on the concept of noninterdependent processes~\cite{asynchronous},  suggests that the observed $a_1$ intrinsically decreases as $v_2$ increases, implying that $\langle a_1\rangle=a_1|_{v_2=0}\cdot(1-\langle v_2 \rangle)$. 
Here, $v_2$ represents single-particle $v_2$, typically on the order of 5\%.
Moreover, higher-order effects could exist even without considering the noninterdependent effect~\cite{1st-ESS}. Hence, to reestablish
the CME signal for the inclusive data, we need to multiply the $y$-intercept of $\Delta\gamma_{112}$ by a  factor of $(1-\langle v_2\rangle)^2$, to the first order.

In the past, various event generators have been employed to investigate the background contributions to the $\gamma_{112}$ correlator.  Notably, models like PYTHIA~\cite{pythia}, HIJING~\cite{hijing1,hijing2}, MEVSIM~\cite{mevsim}, and UrQMD~\cite{urqmd} substantially underestimate the background~\cite{STAR-2,STAR-4}. While the AMPT model~\cite{{ampt_1},ampt_5} is able to reproduce some background contributions comparable to experimental measurements~\cite{Subikash}, even a straightforward ESS application using single 
$v_2$ and single $q_2^2$ can effectively eliminate all the AMPT-generated backgrounds~\cite{ESS-1}. 
The outcomes of our four ESS recipes
from the AMPT simulations
(not shown here) exhibit the same ordering as observed in EBE-AVFD, with the tendency to over-subtract the background.
EBE-AVFD is hitherto the most realistic model for the CME simulations and the only one that necessitates a mixed combination of $v_2$ and $q_2^2$ to completely remove the backgrounds in $\Delta\gamma_{112}$. 
We have also explored augmenting the strength of the LCC effect by half in the EBE-AVFD simulations and confirmed that the conclusions drawn from Figs.~\ref{AVFD_1}, \ref{AVFD_2}, and \ref{AVFD_3} remain unchanged. 
The ultimate validation of the new method hinges on its application to real data, such as Au+Au collisions at 3 GeV, where the QGP (along with the CME) is expected to be absent~\cite{3GeV,3GeV_2,3GeV_3}.

\section{Conclucsion} \label{sec:conclusion}

The experimental search for the CME in heavy-ion collisions remains a challenging task because of the flow related background in the observables.
In this work, we propose an innovative ESS scheme based on a previous work~\cite{ESS-1} so as to effectively suppress the non-CME background  and accurately  extract the CME signal under various experimental scenarios. Our comprehension of background suppression has been advanced in several aspects.
First, we improve the definition of the flow vector by  correcting the normalization with a higher-order term.  This refinement ensures a more reliable ESS projection, especially  for events with high multiplicities.

Second, we extensively investigate one of the primary background sources,  flowing resonances.
With the help of a toy model, we  validate that in the absence of the CME, this background arises from the product of resonance $v_2$ and the correlation between the resonance and its decay daughters.
Moreover, we analytically deduce that resonance $v_2$ incorporates the CME signal, which is further supported by EBE-AVFD simulations.
Thus, the utilization of resonance $v_2$ as an elliptic flow variable in the ESS approach is not recommended, as it leads to a severe over-subtraction of background.
Similarly, according to a recent study~\cite{spin_alignment}, the spin alignment  ($\rho_{00}$) of resonances should not be considered solely as a background contribution but rather as another CME observable.
Extra caution is necessary when developing methods for background control   using variables such as resonance $v_2$, $\rho_{00}$, or even invariant mass.

Third, by  incorporating particle pair  information, we expand the ESS technique to include four distinct
combinations of event shape and elliptic flow variables.
The unmixed recipes exploit the azimuthal angles of either
the same POI or the same pairs to construct both the event shape variable and the elliptic flow variable, generating 
 intrinsic
correlations  that
lead to an under-subtraction
of background.
Conversely, the mixed recipe using pair $v_2$ as the elliptic flow variable (and single $q_2^2$ as the event shape variable)  over-subtracts the background, because pair $v_2$ contains the CME signal, albeit in a  diluted form.
Therefore, the optimal event shape selection strategy
involves utilizing  single $v_2$ in conjunction with pair $q_2^2$.
These findings have been supported by EBE-AVFD calculations.
Additionally, in Appendix~\ref{appendix2}, we have demonstrated 
that using an event shape variable devoid of POI yields much larger uncertainties, both statistically and systematically.

In short, we have pinpointed a technically robust ESS approach, which effectively mitigates the flow-related background in the $\Delta\gamma_{112}$ correlator.
We eagerly anticipate the application of the method outlined in this paper to real-data analyses in the pursuit of the CME in the heavy-ion collisions.

\begin{acknowledgments}{
The authors thank Shuzhe Shi and Jinfeng Liao for providing the EBE-AVFD code and for many fruitful discussions on the CME subject.
We also thank Yufu Lin for generating the EBE-AVFD events.
We are especially grateful to Sergei Voloshin  
for the initial inspiration.
Z. X., B. C., G. W. and H. Z. H. are supported
by the U.S. Department of Energy under Grant No. DE-FG02-88ER40424 and by the National Natural Science Foundation of China under Contract No.1835002.
A. H. T. is supported by the U.S. Department of Energy under Grants No. DE-AC02-98CH10886 and DE-FG02-89ER40531.
}
\end{acknowledgments}

\appendix

\section{Event shape variable exclusive of POI}
\label{appendix2}

\begin{figure}[tb]
\begin{minipage}[c]{0.235\textwidth}
\includegraphics[width=\textwidth]{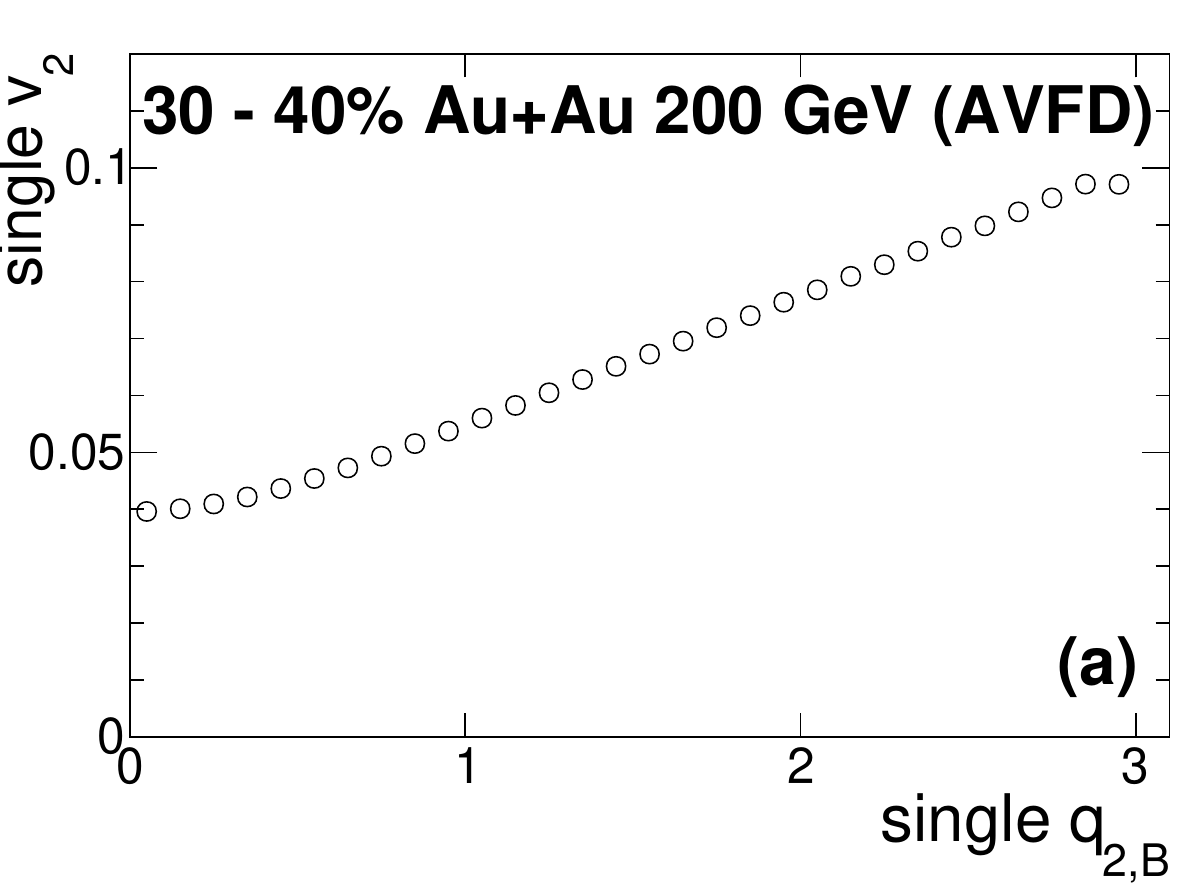}

\end{minipage}
\begin{minipage}[c]{0.235\textwidth}
\includegraphics[width=\textwidth]{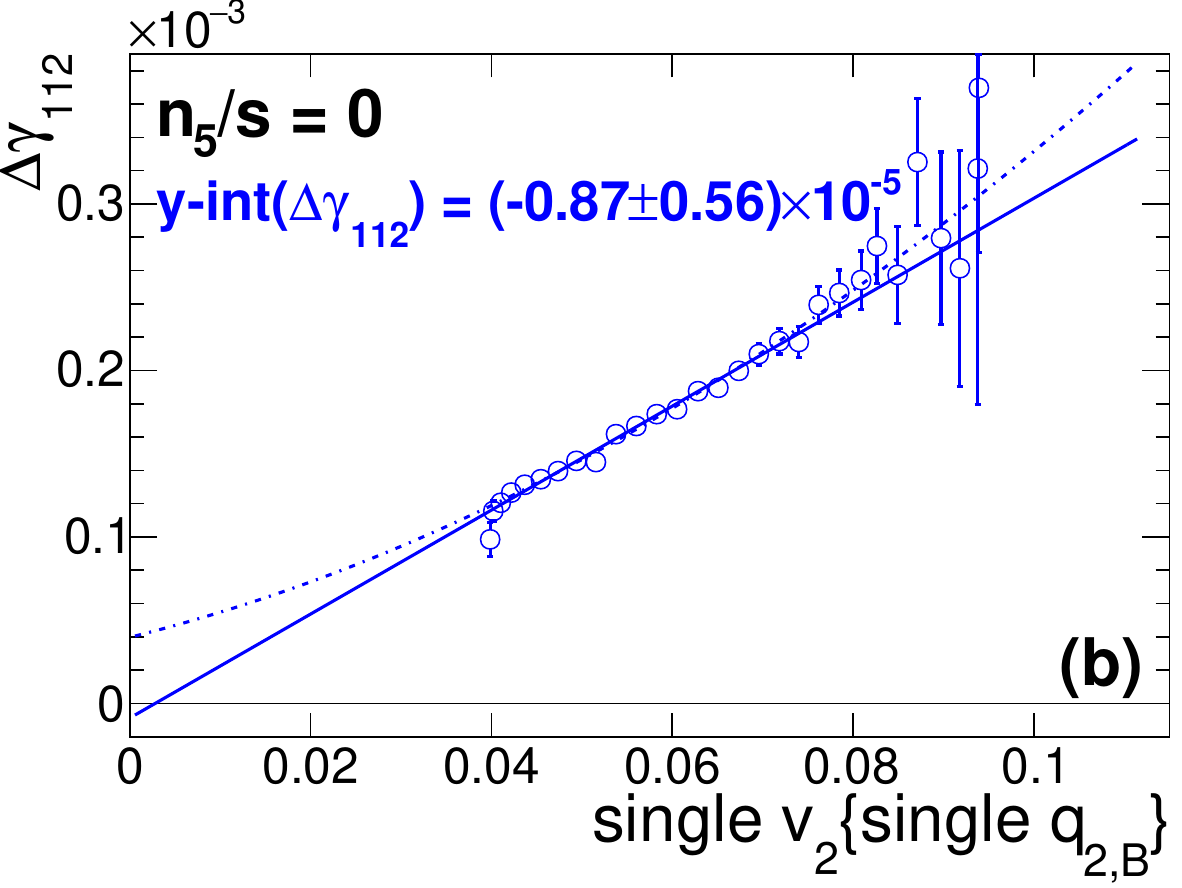}
\end{minipage} 
\begin{minipage}[c]{0.235\textwidth}
\includegraphics[width=\textwidth]{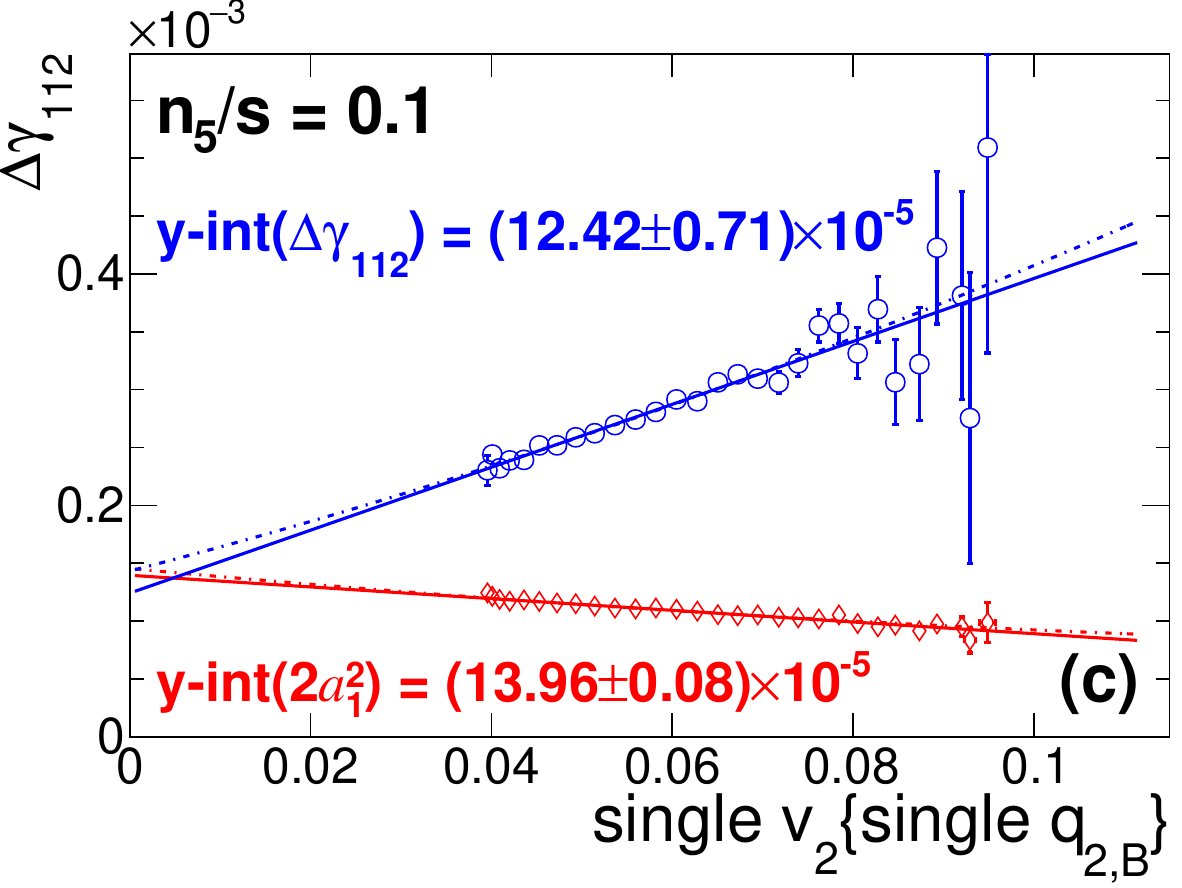}

\end{minipage}
\begin{minipage}[c]{0.235\textwidth}
\includegraphics[width=\textwidth]{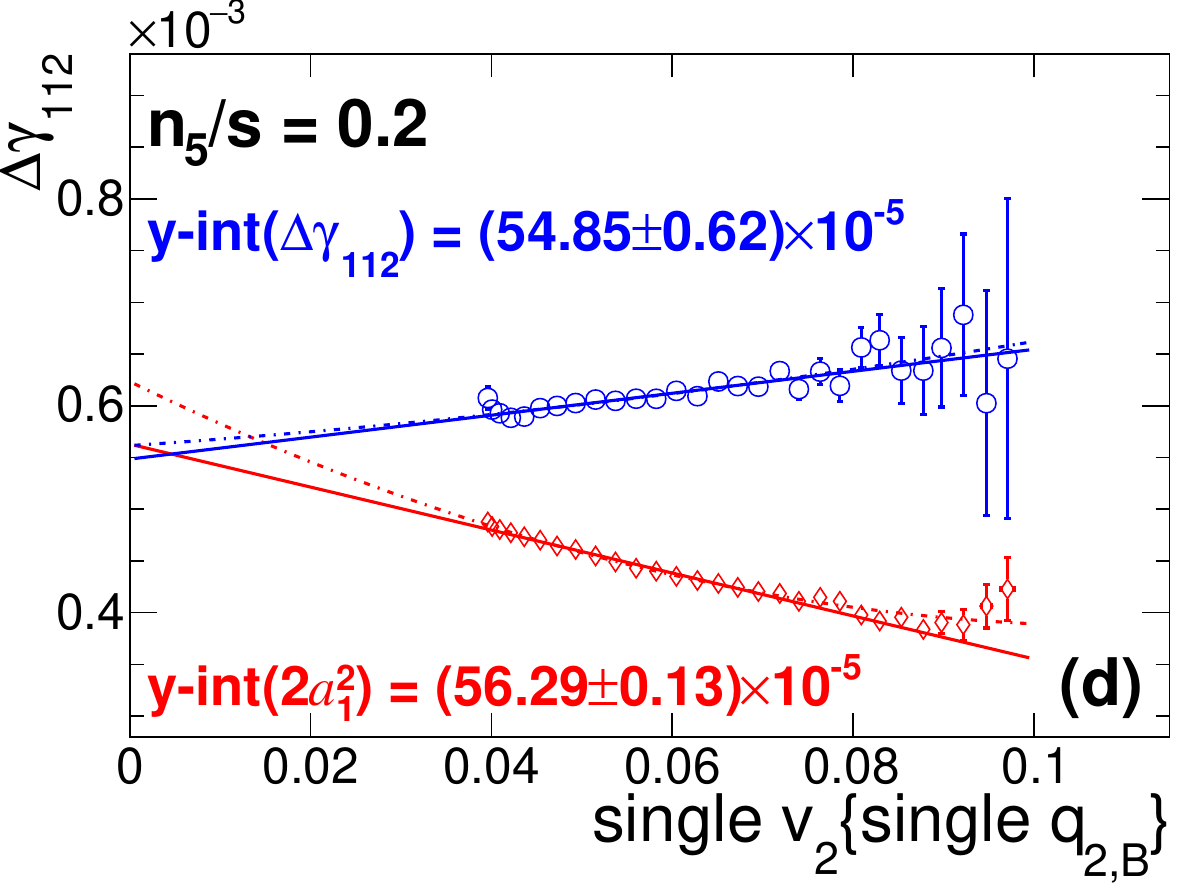}
\end{minipage} 
\caption{EBE-AVFD simulations of single $v_2$ as a function of single $q_{2,{\rm B}}$ (a) and $\Delta\gamma_{112}$ as a function of single $v_2$ categorized by single $q_{2,{\rm B}}$ in the  scenarios of $n_5/s=0$ (b), 0.1 (c), and 0.2 (d) in 30--40\% Au+Au collisions at 200 GeV. The particles of interest are  charged pions. The $y$-intercepts obtained with linear fits (solid lines) reflect the efficacy of these ESS recipes in reducing the elliptic-flow-related background. The second-order polynomial fits (dashes lines) demonstrate the systematic uncertainties due to long extrapolations.
}
\label{AVFD_qB}
\end{figure}

To examine the event shape variable exclusive of POI, we construct $q_{2,{\rm B}}$ as defined in Eq.~(\ref{eq:new_q}) using single particles with $1 < |y|<2$, while POI are within $|y|<1$. 
Figure~\ref{AVFD_qB}(a) shows the EBE-AVFD simulations of single $v_2$ as a function of single $q_{2,{\rm B}}$ in 30--40\% Au+Au collisions at 200 GeV.
Compared with the $q_2^2$ involving POI, $q_{2,{\rm B}}$ exhibits a weaker correlation with single $v_2$, and the latter remains sizable at zero $q_{2,{\rm B}}$.
Figures~\ref{AVFD_qB}(b)--\ref{AVFD_qB}(d) present $\Delta\gamma_{112}$ as a function of single $v_2$ categorized by single $q_{2,{\rm B}}$ in the  scenarios of $n_5/s=0$, 0.1, and 0.2, respectively.
The $y$-intercepts obtained
with linear fits (solid lines) indicate a certain degree of over-subtraction of background.
The extensive blank regions at low $v_2$ considerably amplifies the statistical uncertainties of the $y$-intercepts, leading to an approximate tripling of the uncertainties compared with using event shape variables involving POI. Moreover, if we alternatively apply second-order polynomial fits (dashes lines), the variations of the $y$-intercepts could exceed the statistical uncertainties, exacerbating the issue further.
Conversely,  employing POI for the event shape variable
renders significantly more reliable results, in terms of both statistical accuracy and systematic consistency.

\section{Resonance $v_2$}
\label{appendix1}
For simplicity, we set $\Psi_{\rm RP} = 0$, and assume that particles $\alpha$ and $\beta$ have the same  $p_T$. We thus obtain  $\tan\varphi^{\rm p} = (\sin\varphi_\alpha+\sin\varphi_\beta)/(\cos\varphi_\alpha + \cos \varphi_\beta)$, where $\varphi^{\rm p}$ is the azimuthal angle of this particle pair. Then, the following trigonometric identity becomes
\begin{eqnarray}
\cos 2\varphi^{\rm p} &=& (1-\tan^2\varphi^{\rm p})/(1+\tan^2\varphi^{\rm p})\nonumber\\
&=&\frac{\cos2\varphi_\alpha+\cos2\varphi_\beta+2\cos(\varphi_\alpha+\varphi_\beta)}{2+2\cos(\varphi_\alpha - \varphi_\beta)}.
\label{eq:resonancev2_1}
\end{eqnarray}
Averaging Eq.~(\ref{eq:resonancev2_1}) over OS or SS pairs and over events, we have
\begin{eqnarray}
\langle \cos2\varphi^{\rm OS(SS)}\rangle &\approx& (v_2^{\pi} + \gamma^{\rm OS(SS)})/(1+\delta^{\rm OS(SS)}) \nonumber\\
&\approx& 
v_2^{\pi} + \gamma^{\rm OS(SS)} - v_2^{\pi}\delta^{\rm OS(SS)},
\end{eqnarray}
where $\delta \equiv \langle \cos(\varphi_\alpha-\varphi_\beta) \rangle$ is the two-particle correlation. We have assumed $|\delta| \ll 1$ and ignored higher-order terms.
Resonance $v_2$ is the elliptic flow for the excess of OS over SS pairs,
\begin{eqnarray}
v_2^{\rm res} &=&\frac{N_{\rm{OS}} \langle\cos2\varphi^{\rm{OS}}\rangle - N_{\rm{SS}}\langle\cos2\varphi^{\rm{SS} }\rangle}{N_{\rm{OS}} - N_{\rm{SS}}} \\
&\approx& v_2^{\pi} +\frac{N_{\rm{OS}}+N_{\rm{SS}}}{2(N_{\rm{OS}}-N_{\rm{SS}})}(\Delta\gamma_{112}-v_2^\pi\Delta\delta) \nonumber \\
& & +\frac{1}{2}(\gamma^{\rm OS} +\gamma^{\rm SS}) - \frac{v_2^\pi}{2}(\delta^{\rm OS}+\delta^{\rm SS}),
\label{resonancev2_3}
\end{eqnarray}
where $\Delta\delta = \delta^{\rm{OS}} -\delta^{\rm{SS}}$. Regarding the connection to the CME signal, resonance $v_2$ is notably similar to the signed balance functions~\cite{AVFD5}.

{}
\end{document}